# Cross-species digital xenocortication model delineates molecular correlates of human brain connectome.

Gennadi V. Glinsky[1]

[1] Institute of Engineering in Medicine

University of California, San Diego

9500 Gilman Dr. MC 0435

La Jolla, CA 92093-0435, USA

Correspondence:

genlighttech@gmail.com

## Abstract

In this contribution, the cross-species digital xenocortication (CSDX) model based on a pairing of a human brain structure with one or more mouse brain structures was constructed and implemented to dissect the molecular correlates of the human brain connectome. Design of the CSDX model is based on the assumption that a human brain region and its named mouse counterpart occupy corresponding positions in an evolutionary conserved mammalian brain connectome plan, so that a projection strength measured between two mouse structures is informative about the relation between the two human structures paired with them. In the CSDX model, gene expression for the node was measured in the human brain, while connection strength between two nodes was measured in the mouse brain. CSDX models of different parcellations were employed to explore the potential regulatory associations of brain connectome patterns and constituents of protein-protein interaction (PPI) cascades of hybrid assembly pathways (HAP) of endogenous multiprotein complexes (EMC). Metrics of continuous engagement of the PPI cascade-nucleated protein dimers of CORUM multiprotein complexes have been identified as statistical components of the multivariate transcriptome–connectome associations. It accounts for additional connectome-associated variations beyond the five measured axes in the 82-node CSDX model. Comparisons of the seven-arm models of transcriptome/connectome associations demonstrate that the mouse-derived model generally fits the mouse brain connectome better, while the human-derived model assigns a significantly larger fraction of its transcriptome/connectome association to unique rather than net shared contributions. This study nominates elite panels of top-scoring candidate protein dimers and CORUM multiprotein complexes for follow-up mechanistic studies of their potential roles in structural-functional features governing mammalian brain connectome.

## Introduction

Endogenous multiprotein complexes (EMC) are the essential structural blocks and indispensable functional units of all vital cellular processes across the universe of biological lifeforms. The protein-protein interactions (PPI) between monomer subunits that build the EMC are catalogued in interaction databases that are largely cell-type agnostic. The curated EMC in these databases record sets of proteins that can associate without talking into account whether any particular cell contains all of them or a fraction of them at a defined time point of the lifecycle. This gap between the protein interactome as catalogued and the intracellular protein interactome that is readily available and may contribute to the EMC assembly is the significant problem that potentially undermine most PPI network-based inference in contemporary cell and developmental biology. A model that reasons over the full catalogue of potentially available protein monomers will predict the same complexes everywhere, a conclusion that is not fully compatible with the remarkable diversity of distinct cell types in a multicellular organism. In contrast, a model that reasons over what sub-sets of protein monomers a cell expresses has to specify how expression constrains the EMC assembly.

DNA sequences derived from transposable elements (TEs) provide major contributions to a multitude of structural features and regulatory functions at different levels of genomic and molecular organization of human cells (Lu et al., 2014; Elbarbary et al., 2016; Sundaram and Wysocka, 2020; Hoyt et al., 2022; Rosspopoff et al., 2025). Exceedingly large numbers and immense molecular diversity of ~4.5 million TE-encoded DNA loci comprising approximately 50% of human genome (Hoyt et al., 2022) signify a multitude of formidable challenges for experimental explorations of their biological functions. Results of experimental studies focused on individual TE families facilitated the conceptual advances in understanding the regulatory and pathophysiological features that are common for DNA loci derived from different TE families (Glinsky, 2015-2026; and references therein). Collectively, the experimental observations reported by multiple independent groups of investigators demonstrated the crucial regulatory roles of the transcriptional activation of different families of TEs in human embryogenesis and the early stages of human brain development (Wissing et al., 2012; Gao et al., 2018; Grow et al., 2015; Lu et al., 2024; Guo et al., 2014; Klawitter et al., 2016; Glinsky et al., 2018; Konkel et al., 2020; Parylak and Gage, 2024; Toda et al., 2024; Adami et al., 2025; Fueyo et al., 2025; Glinsky, 2015-2026). These essential for human embryonic development TE-encoded regulatory loci are located in the accessible chromatin regions that became silenced in differentiated cells. However, recent observations added a novel layer of information supporting the concept of important regulatory roles of TE-derived DNA loci residing within transcriptionally silenced heterochromatin regions in execution mechanisms of the broad range of physiological functions related to the homeostasis maintenance of a cell.

Recently, the initiation mechanism of protein-protein interactions (PPI) of hybrid assembly pathways (HAP) of EMC has been reported (Glinsky, 2026a; b). The HAP concept attributes the initiation step of PPI cascade chain reactions to genomic DNA sequences predominantly residing in transcriptionally-silent heterochromatin regions. Chromosome-naïve networks of DNA sequences derived from transposable elements (TE) function as initiation templates of PPI chain reactions which continually create and release from DNA templates cell type-specific pools of hetero- and homodimers. Upon release from DNA templates, protein dimerization products serve as on-demand attractors of specific sets of monomers for the EMC assembly completion. At the HAP initiation phase, a set of ~700 proteins designated initiator transcription factors (TF) bind to double-stranded TE-seeded DNA templates in a sequence-specific manner and upon binding they are poised to initiate PPI cascade chain reactions creating cell type-specific pools of protein homo- and heterodimers (Glinsky, 2026 a; b).

Gene Set Enrichment Analyses (GSEA) demonstrated that this set of initiator TFs could be further divided into 2 families based on distinct patterns of their expression in human brain regions: 1) a set of 545 proteins manifesting region-specific profiles of significantly enriched or depleted expression in human brain; 2) a set of 171 proteins expression of which is detectable in hundreds of brain regions with no evidence of region-specific expression enrichments or depletion (Glinsky, 2026 a;b). Independent follow-up GSEA of sets of 545 and 171 initiator TFs revealed numerous examples of identical significantly enriched records of biological and disease pathways (11 complementation categories of GSEA employing the WikiPathways 2024 Human database); Mammalian Phenotypes (104 complementation categories of GSEA employing the MGI Mammalian Phenotype Level 4 2024 database); phenotypes of human diseases (37 complementation categories of GSEA employing the DisGeNET database). Based on these observations, two families of HAP constituents were designated as the principal initiator TFs and complementary initiator TFs (detailed summaries of these findings have been reported in Glinsky, 2026a; b; c).

Analyses of PPI cascade patterns of initiator TFs demonstrated that HAP-associated PPI chain reactions could concurrently diverge into 2 modes of operations: TF-TFs PPI cascade branch and TF-HUB proteins PPI cascade branch (Glinsky, 2026 b). The sequence of PPI cascade chain reactions reach the PPI cascade equilibrium at saturation levels of PPI constituents of 245 TFs and 351 TF-HUBs of TF-TFs PPI cascade branch and TF-HUB proteins cascade branch, respectively (Glinsky, 2026 b). Therefore, based on these observations 2 mechanistically distinct families of HAP constituents has been identified: 1) a set of proteins comprising the equilibrium (plateau) level constituents of the TF-TFs PPI cascade branch; 2) a set of proteins comprising the equilibrium (plateau) level constituents of the TF-HUB proteins PPI cascade branch. Overall, the canonical panel of proteins engaged in PPI cascade chain reactions during the execution of HAP of EMC on DNA templates consists of 4 distinct families of PPI cascade constituents: 1) principal initiator TFs; 2) complementary initiator TFs; 3) an equilibrium (plateau) stage PPI TF-TFs binders; 4) an equilibrium (plateau) stage PPI TF-HUB protein's binders. One of the limitations of these foundational observations was that the genetic identities and numerical values assigned to each family of PPI cascade HAP constituents were inferred from genome-wide analyses of TFBS and PPI cascade chain reactions associated with LTR5_Hs loci and other TE families. Detailed evaluations of these families of HAP constituents in the contexts of diverse patterns of intracellular protein availabilities in human cells have been reported only recently (Glinsky, 2026c). Within the human body-wide contexts of intracellular proteomes of 154 human cell types, the following key features of the molecular architecture and functional performance of the 4 families of PPI cascade constituents have been identified: a) ubiquitous expression; b) balanced representation; c) universal availability and engagements; d) constitutive {invariant) and cell type-specific assembly profiles of protein dimers of EMCs. Most recent analytical evidence indicate that genome-wide chromosome-naïve networks of transposon-seeded DNA-templates nucleate the initiation of PPI cascade chain reactions with the capacity to facilitate the concurrent iterative creation in human cells of protein dimers for the assembly of both the membraneless biomolecular condensates and the structured multiprotein complexes (Glinsky, 2026).

Several distinct molecular functions attributed to the PPI cascade constituents bound to TE-encoded DNA sequences could facilitate their associations with the brain connectome and potentially affect its structural-functional integrity. Two principally distinct mechanisms of the PPI constituents' actions are conditioned on the chromatin states of the genomic regions in which specific sets of TE-encoded DNA loci are located. TE-encoded DNA loci located in the transcriptionally-silent heterochromatin regions function as DNA templates for initiation of the PPI cascade chains creating the cell type-specific intracellular pools of protein dimers, thus nucleating the protein dimerization reactions for EMC assembly (Glinsky, 2026a; b; c). In contrast, TE-encoded DNA loci residing in genomic regions of the transcriptionally active chromatin state (for example, ATAC-seq-defined regions of chromatin accessibility) execute of wide-range of regulatory functions (Lu et al., 2014; Elbarbary et al., 2016; Sundaram and Wysocka, 2020; Rosspopoff et al., 2025), primarily affecting the expression of specific sets of target genes. In the contexts of transcriptionally active accessible chromatin regions, TE-encoded DNA loci exert their regulatory functions by acting as promoters, enhancers, non-coding RNA molecules, and/or boundary elements of topologically associating domains of 3D structures of the linear chromatin fibers. These two principally distinct mechanisms of the PPI constituents' actions could be viewed as 2 main vectors of their biological functions that are defined by the locations of DNA loci to which the PPI constituents are bound. In turn, the PPI constituents bound to DNA loci located in transcriptionally silent heterochromatin execute their activity through 2 distinct modes of operations with biologically distinct endpoints, namely TF-TF PPI cascade branch and TF-HUB proteins PPI cascade branch. These 2 distinct operational modes could be viewed as secondary vectors of biological functions of the PPI cascade constituents bound to the DNA loci located within transcriptionally silent heterochromatin. Similarly, regulatory actions of the PPI cascade constituents bound to DNA loci located in transcriptionally active genomic regions could be viewed through the lenses of multiple mechanistically-distinct secondary vectors of their biological functions. In a cell operating within a genomic context of ~4.5 million TE-derived DNA loci continuing concurrent execution of these multi-vectored regulatory networks would require the engagement of multi-component panels of regulatory elements assembled from the variable sub-sets of the PPI constituents. Therefore, genomic regulatory networks engaging in their operation the PPI cascade constituents may be regarded as the multi-vectored multi-component biological system designed to support the key elements of the cellular homeostasis.

In this contribution, the potential relevance of the PPI cascade constituents to the molecular correlates of the human brain connectome has been explored. To accomplish this goal, the cross-species digital xenocortication (CSDX) model was constructed and implemented. The foundation of the architectural design of the CSDX model was built on a pairing of a human brain structure with one or more mouse brain structures. The design idea of the CSDX model is based on the assumption that a human brain region and its named mouse counterpart occupy corresponding positions in an evolutionary conserved mammalian brain connectome plan, implying that a projection strength measured between two mouse structures is informative about the connectome between the two human structures paired with them. In the CSDX model, gene expression for the node was measured in the human brain, while connection strength between two nodes was measured in the mouse brain. While this project was in advanced execution stages, the pioneering breakthrough study was published describing the experimental model of developmental xenocortication using human-derived organoids in mice (Kaganovsky et al., 2026). In this work, a transplantation platform was established using a genetic strategy to deplete glutamatergic neurons from mouse neocortex and hippocampus and neonatally engraft the mouse cortical cavity with human stem-cell-derived cortical organoids. In this experimental xenocortication model, human cells appear efficiently engrafted in developing mouse brain with human cells occupying most of the cortical volume and generating an apparent diversity of functional human cortical cell types, including layer 5 extratelencephalic projection neurons (Kaganovsky et al., 2026). It has been documented that human cortical neurons structurally and functionally integrate in vivo with the mouse nervous system in the viable functional xenocortical adult mice with broadly preserved locomotion (Kaganovsky et al., 2026), thus providing direct experimental support for the validity of the idea of cross-species digital xenocortication model developed and implemented in this contribution.

## Results

### Design of the digital cross-species xenocortication model for identification of the molecular correlates of the human brain connectivity.

Designed and implemented in this study digital cross-species xenocortication (CSDX) model is based on a pairing of a human brain structure with one or more mouse brain structures. In the CSDX model, gene expression for the node is measured in the human brain, while connection strength between two nodes is measured in the mouse brain. No human connectivity data enters the analysis, and no mouse expression data employed in the model. In the CSDX model, the parcellation of the brain nodes was the analytical instrument that allowed the gene expression and the connection strength be read against each other. Therefore, all reported results of the analyses are based on the assumption that a human brain region and its named mouse counterpart occupy corresponding positions in a conserved mammalian brain connectome plan, so that a projection strength measured between two mouse structures is informative about the relation between the two human structures paired with them. The instances where that assumption is weak are explicitly disclosed, per node, in the homology grade description (Methods). The instances where the crosspiece's pairing assumption fails altogether were excluded rather than pair the human structure with a poor substitute (Methods; section S5.4).

The construction of the model is designed as a transfer of the mouse connectome data onto human brain regions. The three parcellations employed in this study are three resolutions of the same transfer: 73 nodes (the base map), 92 nodes (the same map with ten parents split as finely as both human and mouse atlases allow), and 82 nodes (the 92 node parcellation restricted to the brain node pairings whose homology call rests on a named correspondence rather than on relative position). The complexity of the pairing human and mouse brain nodes is illustrated by the example where the lobe-level human panels (21 panels parcellated by lobe) were paired against the Allen Mouse Brain Connectivity Atlas. The mouse isocortex in the Allen Common Coordinate Framework (CCF) is parcellated by area, not by lobe. Matching the two produced the connectivity data for 13 of the 21 panels and nothing for the other eight. The misses were restricted to Cerebral Cortex; Frontal Lobe; Parietal Lobe; Temporal Lobe; Occipital Lobe; Limbic Lobe; Insula, indicating that every miss is cortical and every match is subcortical. The human-mouse pairing gap is structural: the mouse atlas has no object at the granularity of a human lobe, so no amount of additional lobar sampling can close it. The only route to a cortex-inclusive test is to decompose the human cortex into areas that the mouse atlas also names. That approach, the area-level nodes, not lobes, is what this model employs. Consequently, the one lobe-level result specific to the hybrid-assembly-pathway (HAP) protein families analyzed in this study was a six-lobe gap in the P545 family ($z = +2.51$). Detailed descriptions of data sources and exact retrieval and data processing protocols are reported in the Methods section of the study.

**Per-node availability analyses of distinct HAP families of the PPI constituents.**

The CSDX model base map of 73 nodes was selected from a set of 109 candidate nodes (homology tier A or B; Methods) by requiring each of 73 nodes to clear two detection floors: >= 12 samples and >= 4 of 6 donors (Figure 1). The 73 nodes represent 10 brain regions, ranging from 13 isocortex nodes to 3 nodes each of amigdala and hypothalamus (Figure 1). Thus, the model parcellation is subcortically weighted: 13 isocortical nodes against 60 non-isocortical. This setting is a consequence of AHBA sampling being dense in the brainstem and cerebellum and the mouse atlas naming those structures unambiguously. In contrast, the olfactory areas contributed no nodes at all because they failed to clear the sample selection floor. Therefore, the olfactory connectivity has not been analyzed in the CSDX model setting, including the reported below results of the analyses of the rostrocaudal (anterior-posterior) gradient, which starts at frontal cortex rather than at the olfactory bulb. A node enters the parcellation only if it has ≥12 samples from ≥4 donors, so donor coverage is already carried by the node admission rule and gene detection threshold was defined at 50%: a gene is present in a node when more than half that node's samples call it present (based on the AHBA PACall flag calls identifying the gene present; Methods).

Per-node availability analyses of the four HAP families of the PPI cascade constituents were carried-out under three detection-axis rules, at three parcellations (Figures 1 and 2). Similar to the findings of the investigation of the constituents of PPI cascades elucidated within the context of cell type-specific intracellular protein monomer pools documented in the Human Protein Atlas (HPA) of 154 human cell types (Karlsson et al., 2021; Glinsky, 2026), per-node availability analyses revealed the common key features of the molecular architecture of the 4 families of PPI cascade constituents within different CSDX model contexts of intracellular proteomes: a) ubiquitous expression; b) balanced representation; c) universal availability and engagements.

Notably, 256 of 540 (47%) measured members of the principal initiator TFs (P545) family are detected in all 73 nodes (255 of 540 are expressed in all 82 nodes and 92 nodes), and 193 of 239 (81%) members of the TF-HUB proteins (H350) family are ubiquitously expressed. In contrast, only relatively small fractions of proteins comprising the PPI cascade constituents were never detected in the CSDX model: 76; 21; 17; and 5 members of the P545; C171; T245; and H350 families, respectively. Members of the H350 family are detected more than the control background set everywhere: on the detection rule, significant detection values were observed at every one of the 82 nodes; 73 nodes; and 92 nodes. Only 5 of its 239 (2%) measured members are never detected anywhere in the contexts of the CSDX model (BTK, ESR2, IKBKE, TNFRSF1B, TP53). In contrast, 76 of 540 (14%) measured members of the P545 family are never detected in any node, the largest never-detected fraction of any family.

Observations of the broad consistent expression of the PPI cascade constituents of the TF-HUB proteins' PPI cascade branch (H350 family) in all analyzed herein human brain nodes of the CSDX model are highly congruent with their roles in synaptogenesis and structural-functional integrity of synapses (Glinsky, 2026). These findings are in agreement with the concept that one of the main biological functions of the PPI cascade constituents of the TF-HUB proteins' PPI cascade branch (H350 family) is to supply the protein dimers for the assembly of multiprotein complexes of synapses, in particular, the key elements of the post-synaptic density (PSD), thus fulfilling one of the essential structural requirements for the establishment of intercellular connectivity networks.

**Principal initiator TFs family is a winner of detection-axis Mantel tests against the connectome.** Node-by-node Jaccard similarity of the family's detection patterns were correlated (Spearman) against the mouse-connectivity edge weights over the same nodes (Table 1; Figure 3; Methods).

**Table 1.** The summary statistics of detection-axis Mantel test against the connectome.

| Family | ρ (82) | p | partialDist | p | partialPanels | p | size-matched draw z |
|---|---|---|---|---|---|---|---|
| **P545** | 0.377 | 0.0002 | 0.230 | 0.0002 | **0.231** | **0.0002** | +2.74 |
| **C171** | 0.217 | 0.0004 | 0.173 | 0.0002 | 0.003 | 0.94 | −0.64 |
| **T245** | 0.325 | 0.0002 | 0.232 | 0.0002 | 0.105 | 0.0016 | +1.34 |
| **H350** | 0.254 | 0.0004 | 0.211 | 0.0002 | 0.039 | 0.39 | −0.19 |
| **init716** | 0.365 | 0.0002 | 0.224 | 0.0002 | 0.226 | 0.0002 | +2.46 |
| **union1122** | 0.377 | 0.0002 | 0.234 | 0.0002 | 0.247 | 0.0002 | +3.02 |

Three distinct statistical estimates of significance were employed in these analyses: raw p from 4,999 joint node-label permutations; partialDist test based on removal of inter-node distance; and partialPanels test that removes the whole-panel detection identity, which is considered the strictest of the three approaches, because it asks whether the family adds anything over the detection pattern of every measured gene at once. Every family correlates with the connectome (Figure 3) based on the raw p values and after inter-node distance removal. Only P545 family (and the two supersets of 716 initiator TFs and the union of 1,122 HAP constituents that contain it) survives removing the whole-panel identity. Analyses of 73 nodes and 92 nodes reproduce this observation exactly (P545 partialPanels 0.257 / 0.214, p 0.0002 both; C171 −0.001 / −0.013, p = 0.97 / 0.68; H350 0.044 / 0.018, p = 0.39 / 0.67).

**The breadth-matched null analysis confirms the superior performance of the principal initiator TFs family.**

The size-matched draw analysis in the previous section has a detection level limitation: 60.1% of random measured genes are detected in all 82 nodes against 47.2% of P545, and a family with a larger undetected fraction has more structure in its Jaccard matrix before the anatomy impact enters. The breadth-matched null analysis removes this limitation: each member is replaced by a non-member detected in exactly the same number of nodes (nearest stratum on fallback), so the drawn set carries the family's breadth profile gene for gene and the only thing left to differ is which nodes. Results of these analyses in which the exact-breadth matching was achieved for 100% of members at all three parcellations are reported in the Table 2.

**Table 2.** The statistical summary of the breadth-matched null analysis.

| Family | universal % | ρ | null mean ± sd (82) | z 73 / 82 / 92 | p (82) |
|---|---|---|---|---|---|
| **P545** | 47.2 | 0.377 | 0.274 ± 0.024 | +4.60 / **+4.22** / +4.01 | 0.0050 |
| **C171** | 60.4 | 0.217 | 0.237 ± 0.042 | −0.48 / −0.47 / −0.50 | 0.67 |
| **T245** | 66.8 | 0.325 | 0.269 ± 0.041 | +1.73 / +1.36 / +1.15 | 0.085 |
| **H350** | 80.8 | 0.254 | 0.266 ± 0.047 | −0.05 / −0.27 / −0.41 | 0.63 |
| **init716** | 50.1 | 0.36 | 0.281 ± 0.020 | +3.88 / +4.19 / | 0.00 |
| | | 5 | | +3.67 | 50 |
| **union1122** | 59.6 | 0.377 | 0.296 ± 0.021 | +4.31 / +3.82 / +3.66 | 0.0050 |

Legend: At 200 draws, seed 9021; the empirical p floor is considered 1/201 = 0.0050.

Members of the P545 family's performance margin over the connectome is about where its members are detected, not how broadly they are expressed. Reported in the Table 1 raw correlations of the C171 and H350 families appear fully explained by their breadth profiles (compare the p values reported in the Table 1 and Table 2). Based on these findings, it has been concluded that only P545 family (and the two supersets of 716 initiator TFs and the union of 1,122 HAP constituents that contain it) survives the breadth-matched null analysis.

**Anatomical division separation analysis: members of the principal initiator TFs' family (P545) differentiate human anatomical divisions and tracks the connectome.**

Statistical approach of this analyses was mean between-division Jaccard dissimilarity minus mean within-division dissimilarity over node pairs, scaled by the overall mean; z estimated against the same breadth-matched null. In this analysis, positive z indicates that the family's detection pattern separates the anatomical divisions (isocortex, hippocampus, amygdala, cerebral nuclei, thalamus, hypothalamus, midbrain, pons, medulla, cerebellum) better than matched genes do.

Reported in this section of the Results observations are supported by the statistic that was built to test the claim that a biological function of the Principal initiator TFs is to differentiate brain regions from one another. It can fail, and for three of the six families it does (Table 3). The dissociation between the different PPI cascade families appears clean: P545 family of the principal initiator TFs separates divisions and tracks the connectome after the whole-panel identity is removed. In contrast, C171 family does neither and H350 family does neither. Since init716 family is the sum of members of P545 and C171 families, the 716 initiator panel's statistical significance signal is carried by its Principal initiator TF members of P545 family.

**Table 3.** Statistical summary of the anatomical precision analyses.

| Family | z 73 | z 82 | z 92 | p (82) |
|---|---|---|---|---|
| **P545** | **+3.98** | **+3.03** | **+2.13** | **0.0050** |
| **C171** | −0.02 | −0.92 | −0.76 | 0.82 |
| **T245** | +0.87 | +0.85 | +0.60 | 0.22 |
| **H350** | −0.95 | −0.70 | −0.10 | 0.79 |
| **init716** | +3.45 | +2.55 | +1.42 | 0.0050 |
| **union1122** | +3.34 | +2.54 | +1.89 | 0.0050 |

It has been noted that the z values appear to fall as the parcellation gets finer (+3.98 → +3.03 → +2.13). Additional tests were carried-out to determine whether the cause of this is the finer parcellation or the remapping precision (Table 4). It has been determined that the remapping precision is the cause: restricting to precisely remapped nodes raises P545 z values to +6.10 at 82 nodes and to +5.79 at 92 nodes.

**Table 4.** Statistical summary of the precision remapping analysis.

| Arm | nodes | P545 divSep z | C171 | T245 | H350 | init716 | union1122 |
|---|---|---|---|---|---|---|---|
| **82 all** | 82 | +3.03 | −0.92 | +0.85 | −0.70 | +2.55 | +2.54 |
| 82 precise | 61 | **+6.10** | +1.49 | +1.55 | −0.64 | +5.94 | +5.21 |
| 82 confidence A | 53 | **+4.55** | +0.33 | +1.30 | −0.04 | +4.29 | +4.05 |
| **92 all** | 92 | +2.13 | −0.76 | +0.60 | −0.10 | +1.42 | +1.89 |
| 92 precise | 63 | **+5.79** | +1.29 | +1.59 | −0.63 | +5.49 | +4.78 |
| **92 imprecise** | 29 | +2.35 | −1.48 | −0.56 | +0.74 | +1.74 | +1.62 |

Dropping 21 loosely mapped nodes from the 82-node set doubles the statistic (+3.03 → +6.10) on fewer node pairs, which is the opposite of what the dilution of the node numbers predicts. C171 family moves from −0.92 to +1.49 and stays short of significance; H350 family does not move at all.

**Transcriptional correlates of the human brain connectome.**

The PPI cascade HAP constituents that are expressed in a cell should be regarded as integral components of the intracellular transcriptome and proteome. Overall, transcriptional identity tracks measured projection strength across the 2,628 pairs at $\rho = +0.261$, $p = 0.0002$, and it survives the control for a spatial comparison: partialling inter-centroid distance leaves +0.202 at the same p value (Figure 4; panel B1). Thus, brain areas that project to each other look alike transcriptionally, and not only because they are near each other. In addition to the global transcriptome-connectome tracking, transcriptomic analyses were repeated four times and reported in each panel as a hexbin of the 2,628 area pairs (Figure 4): the rank of transcriptional identity is plotted on x, the rank of measured brain regions projection strength is plotted on y, a red least-squares line through the ranks, and a colour scale counting pairs per hexagon. The y axis is identical in all five transcriptome analysis panels. What changes is the gene set the identity of which is restricted to before the overlap is measured — the 1,122-protein HAP union (A1; $\rho = +0.229$), P545 (A2; $\rho = +0.256$), T245 (A3; $\rho = +0.180$), H350 (A4; $\rho = +0.230$). Each area's identity is its 4,000-gene marker panel intersected with that family, and a pair's identity is the size of the intersection of the two restricted panels as a percentage of the smaller one. The box in each panel reports three Mantel correlations against the connectome: raw, with inter-centroid distance partialled out, and with the whole-panel identity matrix partialled out. The grey footnote in each panel reports the family's coverage (how many of its proteins fall in some panel), the median and range of restricted panel size per area, and $\rho_W$, the Spearman correlation between that panel's identity matrix and the whole-panel one. Numerical p value is from 4,999 joint node-label permutations (all $p = 0.0002$, the permutation floor). Panel B4 illustrates the strongest PPI cascade family claim stating that reached plateau recovers the ten divisions in every family, the values of the Adjusted Rand Index (ARI) versus anatomical division: 0.3209 (P545) to 0.4720 (T245) at z 17.3 to 24.0, all $p = 0.0005$, the permutation floor (Figure 4; panel B4). While no single transcriptome subset reproduces the full transcriptome panel, several separately defined axes each beat their rank-matched controls, and the relation is nearly unharmed by deleting any one gene class. That combination of features (several redundant, distributed contributing axes) is recapitulate features of what a multifactorial identity signal looks like. The relation between the transcriptome and connectome is multifactorial in the sense that no single gene expression property accounts for it in full and several separately defined properties each recover the statistically-significant relation, but the properties do not appear additive. Two contributions appear separable: how steeply a gene expression varies along the rostro-caudal axis, and whether it is a HAP constituent.

**Delineation of multi-dimensional associations of the brain connectome and the multi-vectored regulatory functions of the PPI cascade constituents.**

Genomic regulatory networks engaging in their operation the PPI cascade constituents may be regarded as the multi-vectored multi-component biological system designed to support the key elements of the cellular homeostasis (Introduction). Results of the present analyses are in agreement with the hypothesis that activities of PPI cascade constituents may contribute to the molecular architecture supporting the foundation of the human brain connectome. Among the PPI cascade constituents, the principal initiator TFs (P545) manifest most variable expression patterns across human brain nodes; it separates divisions and tracks the connectome after the whole-panel identity is removed. The constituents of the TF-HUB proteins' PPI cascade branch (H350 family) manifest the broad consistent expression in all analyzed herein human brain nodes of the CSDX model, the property that is highly consistent with their proposed roles in synaptogenesis and structural-functional integrity of synapses (Glinsky, 2026). It was of interest to investigate whether other molecular functions attributed to the PPI cascade constituents bound to TE-encoded DNA sequences could reflect their associations with the brain connectome and support the inference of the potential effects of the PPI cascade constituents on the brain connectome structural-functional integrity.

**Table 5.** Every family, in both libraries, tracks projection strength at almost exactly the whole-transcriptome value.

| library | family | TFs used | Mantel | p | ρ with whole panel | partial \| whole panel | p |
|---|---|---|---|---|---|---|---|
| **JASPAR** | all profiled TFs | 630 | +0.2672 | 0.0002 | 0.9663 | +0.0669 | 0.0458 |
| **JASPAR** | P545 | 452 | +0.2661 | 0.0002 | 0.9666 | +0.0555 | 0.0962 |
| **JASPAR** | C171 | 145 | +0.2660 | 0.0002 | 0.9776 | +0.0411 | 0.1914 |
| **JASPAR** | T245 | 132 | +0.2617 | 0.0002 | 0.9813 | +0.0341 | 0.2662 |
| **JASPAR** | H350 | 16 | +0.2504 | 0.0002 | 0.9879 | −0.0488 | 0.0864 |
| **JASPAR** | union1122 | 599 | +0.2664 | 0.0002 | 0.9672 | +0.0625 | 0.0608 |
| **JASPAR** | init716 | 597 | +0.2665 | 0.0002 | 0.9665 | +0.0630 | 0.0590 |
| **ChEA** | all profiled TFs | 345 | +0.2657 | 0.0002 | 0.9678 | +0.0347 | 0.2510 |
| **ChEA** | P545 | 146 | +0.2611 | 0.0002 | 0.9620 | +0.0242 | 0.4212 |
| **ChEA** | C171 | 34 | +0.2650 | 0.0002 | 0.9717 | +0.0404 | 0.1634 |
| **ChEA** | T245 | 161 | +0.2665 | 0.0002 | 0.9665 | +0.0417 | 0.1646 |
| **ChEA** | H350 | 26 | +0.2757 | 0.0002 | 0.9634 | +0.0777 | 0.0084 |
| **ChEA** | union1122 | 234 | +0.2596 | 0.0002 | 0.9673 | +0.0047 | 0.8760 |
| **ChEA** | init716 | 180 | +0.2587 | 0.0002 | 0.9624 | +0.0115 | 0.7024 |

Transcription factors constitute a majority of the PPI cascade constituents, which when bound to DNA loci located in the open, transcriptionally-active chromatin regions regulate the target gene expression and their contribution to regional transcriptional identity may be tracked in its targets' transcript expression levels. Next, the relations between the projection strengths and expression levels of genes comprising the candidate regulatory targets of HAP constituents was examined. Results of these analyses demonstrated that every PPI cascade family, in both independently identified libraries of its target genes, tracks projection strength at almost exactly the whole-transcriptome level (Table 5; Figure 4). Transcriptome identity built from genes the HAP constituents regulate tracks brain projection strength at nearly exact the whole-transcriptome level: every family, both libraries, ρ = +0.250 to +0.276 (Table 5), all p = 0.0002, all surviving distance partialling. Distance partialling removes about a fifth and changes nothing qualitative (+0.1910 to +0.2237, all $p = 0.0002$). The PPI cascade HAP constituents-restricted identity correlated 0.82–0.96 with the whole-panel matrix; while HAP constituents' target-restricted identity correlates 0.962–0.988.

Based on results of these analyses, it has been concluded that transcriptional identity computed from the predicted targets of the PPI cascade HAP families tracks measured brain projection strength as strongly as whole-transcriptome identity (ρ = +0.25 to +0.28, all $p = 0.0002$, distance-independent), but appear to carry no significant signal once whole-panel identity is partialled out. The spread across all fourteen target panels is +0.2504 to +0.2757 — a range of 0.025, and a whole-transcriptome value of +0.2607 is sitting in the middle of the range.

**Variance decomposition of the five predictors of the commonality analyses at three parcellations of the CSDX model.**

To dissect the unique and shared contributions of the individual predictors of the brain connectivity, variance decomposition analyses were performed (Methods). Analyses of the 73 nodes CSDX model revealed that the unique components sum to 40.7% of it, so most of what the model explains is shared (Figure 5). Two predictors own most significant unique shares: steep rostro-caudal (anterior-posterior) gradient (17.07%, $p = 0.0002$) and HAP constituency (12.71%, $p = 0.0006$). Contributions of breadth 21–40 (6.08%, $p = 0.066$) and distance (4.86%, $p = 0.102$) do not reach significance, and low abundance owns nothing at all (0.01%, $p = 0.9436$). The model collinearity is in the matrices: the five similarity matrices correlate +0.68 to +0.84 over the 2,628 pairs while the gene-level axes correlate $|\rho| \leq 0.26$.

Removing the ten weakest homology calls in the 82 nodes CSDX model (compared to the 92 nodes CSDX model) does not change the shape of the apportionment of the unique and shared components (Figure 5). Model $R^2$ is 0.1508 by fraction and 0.1480 by count, against 0.1672 at 73 nodes and 0.1793 at 92, and unique components sum to 29.3% and 27.8% of it. The shared fraction therefore runs 59.3% at 73 nodes, 70.7–72.2% at 82 nodes and 81.5% at 92 nodes. It seems to track the pair count (2,628 / 3,321 / 4,186) rather than the homology grade of the nodes.

The rostro-caudal (anterior-posterior) axis is the top predictor with a significant unique share in every arm at every resolution (10.13%, $p = 0.0008$ by fraction; 7.22%, $p = 0.0068$ by count). The 82- and 92-node fraction arms make the same three predictors significant — R, D and H (Figure 5). Contribution of HAP constituency appears to decline monotonically across the three arms, 12.71% → 6.66% → 2.87% of the model, and it remains significant at all three parcellations. In the 82-node continuous-profile model, the unique HAP-constituent contribution changes from 0.01004 (6.66% of baseline $R^2$) to 0.01972 (12.02% of expanded $R^2$). The unique rostrocaudal (anterior-posterior) contribution decreased from 0.01528 to 0.01342; while distance predictor contribution increased from 0.01210 to 0.01555.

Notably, P545 family z value falls as the parcellation gets finer (+3.98 at 73 nodes, +3.03 at 82 nodes, +2.13 at 92 nodes), and the possible interpretation is that more nodes dilute the statistic. However, every restriction to better-mapped nodes raises it instead, on fewer node pairs — +6.10 on 61 nodes at 82, +5.79 on 63 at 92 — which is the opposite of what dilution predicts (Figure 6). Definitions of “precision” and “confidence” refer to different mapping criteria of the CSDX model (Methods). High confidence is the 82-node map's own A/B grade (53 nodes; Methods); precise mapping means one AHBA structure group was mapped to one mouse target undivided (61 nodes; Methods). The term “imprecise” designates the 29 nodes built by dividing a parent structure group. In accordance with the PPI cascade family compositions, P545, init716 and union1122 families move together, and the initiator TFs (init716) is the sum of the principal initiator TFs (P545) the complementary initiator TFs (C171). The initiator TF panel's signal is carried by its principal initiator TFs (P545); C171 stays short of significance in all six arms, reaching +1.49 at its highest. H350 family does not move on either axis (−0.70 → −0.04 → −0.64 and +0.74 → −0.10 → −0.63), which is the control the analysis needs: it indicates that remap precision is not simply raising every HAP family's z.

Overall, results of the decomposition analyses support the concept of the model multifactoriality of the transcriptomics vector of relations between gene expression profiles and brain connectome: the shared block remains the majority of the relation in every analyzed arm, and no single predictor’s unique contribution approaches owning the relation.

**Analysis of the relations of node-specific dimer repertoires and CORUM complex occupancy and brain connectivity.**

Analyses of node-specific dimer repertoires and CORUM complex was run over 82 anatomical nodes of the A/B-homology map; each node's expression panel is the top 4,000 genes selected by across-node z-score (Figure 6; Tables 6a and 6b). Each node’s expression panel seeds a hypergeometric PPI cascade over the Enrichr Transcription_Factor_PPIs and PPI_Hub_Proteins libraries (Benjamini-Hochberg q < 0.05, iterated to convergence). One of the analyzed end-points was division means of the number of genes from each of the four HAP families — P545 principal initiator TFs, C171 complementary initiator TFs, T245 TF-TF PPI cascade branch targets, H350 TF-HUB proteins PPI cascade branch genes — falling inside each node's panel. Reported in the Figure 6 (panel B1) divisions are ordered by mean cascade reach, fewest at the bottom, and the four families partially overlap rather than partition the 1,122-gene union of the PPI cascade constituents. Parenthetical numbers report nodes per division. Relative contributions of the PPI cascade families are clearly distinguishable (Figure 6; panel B2). This observation is clearly demonstrated by plotting the size of each node's dimer inventory against its TF-HUB proteins PPI cascade branch H350 family count (purple, Spearman ρ = +0.73) and against its principal-initiator TF P545 family count (red, ρ = −0.42; lines are least-squares fits). Content of the H350 family sets how large a cascade a node can run, while P545 principal-initiator family content varies in the opposite direction.

Two PPI partner protein monomer pools were explored separately throughout the analyses: panel dimers — the partner must be in the node's own top-4,000 gene panel. Strictly node-specific; the primary set;

CNS dimers — the partner may be any of the 8,695 CNS-expressed genes, the declared monomer source (Table 6a). This permits a node-specific hub recruit a monomer the node does not itself mark.

**Table 6a.** Detection of protein dimers summary statistics.

| Quantity | mean | min–max |
|---|---|---|
| **panel dimers** | 1,463 | 0–3,783 |
| **CNS dimers** | 6,888 | 0–14,505 |
| **of those, engaged (both partners cascade products)** | 1,337 | — |
| **CNS-only (partner not a cascade product)** | 5,552 | — |
| **distinct partners** | 3,133 | — |

Legend: panel Dimers = panel Engaged in 82 of 82 nodes, by construction. A panel partner of a reached term is itself in reached by definition, so the engaged flag is vacuous for the panel pool. It is informative for the CNS pool, where 19% of dimers are engaged.

The protein dimers appearing in the most nodes' top-40 panels are: CAMK2A–PRKCB (18 nodes), CAMK2A–GRIN2B (14), CALM1–CAMK2A (13), CAMK2A–SNCA (13), CALM1–CSNK2A2 (13), CAMK2A–CREB1 (12), AR–MAPK3 (12), CAMK2A–DLG4 (12). Homodimers are 0.1% of the top protein dimer lists. Detailed reports of the analyses of protein dimers and CORUM multiprotein complexes are available in the Supplementary Materials.

Perhaps, most interesting observations are illustrated by the rank correlation between the brain connectome and seven node-by-node similarity matrices (Figure 6; panel B3), computed over the 3,321 upper-triangle node pairs. Dark bars give the Spearman correlation coefficient, gold bars report the same coefficient after partialling on inter-centroid Euclidean distance (distance alone: $\rho = -0.209$). In each four-bar setting, solid bars report values obtained from the analyses of all 82 nodes and pale bars report data from the analyses of the 77 nodes whose cascade returns at least one term. Reported significance is from 4,999 joint node-label permutations, with "ns" marking $p \geq 0.05$. CORUM complexes ($n = 1,132$ with ≥ 3 subunits) were tested against the 8,695-gene CNS-expressed background, which also supplies the monomer pool for the CNS-partner dimer arm. All seven arms derive from the same marker panels and the comparison shown is of what survives the panel → cascade → dimer → complex chain.

A continuous CORUM engaged-subunit-fraction profile adds 0.01317 to model $R^2$ beyond the five predictors in Figure 5B (0.15084 → 0.16401) across all 82 nodes (Table 6b). This is 8.03% of the expanded model's explained rank variance; the observed gain has $p = 0.0164$ under 4,999 joint node-label permutations of the outcome. An approximate residual-label sensitivity test, which keeps the five-predictor fit fixed, gives $p = 0.0156$. With five no-reach nodes omitted, the 77-node baseline rises from 0.15533 to 0.18400 ($\Delta R^2 = 0.02867$; 15.58% of the expanded model; $p = 0.0004$, residual-label $p = 0.0006$). The 82-node baseline reproduces Figure 5B's $R^2 = 0.1508$ and its individual unique contributions.

**Table 6b.** Node-specific dimer repertoires and CORUM complex occupancy both track interregional connectivity, and the complex-level profile tracks it most closely.

| Node scope | Added CORUM predictor | Baseline $R^2$ | Expanded $R^2$ | $\Delta R^2$ (unique) | Permutation p |
|---|---|---|---|---|---|
| **All 82** | Continuous engaged-fraction profile | 0.15084 | 0.16401 | 0.01317 | 0.0164 |
| **All 82** | ≥50% engaged complex sets | 0.15084 | 0.15091 | 0.00007 | 0.8838 |
| **All 82** | BH q < 0.05 enriched complex sets | 0.15084 | 0.15238 | 0.00154 | 0.3106 |
| **77 with nonempty cascade** | Continuous engaged-fraction profile | 0.15533 | 0.18400 | 0.02867 | 0.0004 |
| **77 with nonempty cascade** | ≥50% engaged complex sets | 0.15533 | 0.15535 | 0.00001 | 0.9368 |
| **77 with nonempty cascade** | BH q < 0.05 enriched complex sets | 0.15533 | 0.15695 | 0.00162 | 0.3284 |

A total of 1,132 CORUM complexes with ≥ 3 subunits were identified on the background of 8,695 CNS-expressed genes. Median 1 complex per node at q < 0.05 (range 0–59) were detected. A total of 264 distinct multiprotein complexes are scored as significant; none in all 82 nodes; 19 nodes have ≥ 10 CORUM complexes. The complexes that clear BH across the 1,132 tests most often are large housekeeping multiprotein machines: NOP56p-associated pre-rRNA complex (14 nodes), cytoplasmic ribosome (12), profilin 2 complex (9), large Drosha complex (9), LARC complex (9), spliceosome complex (9), DGCR8 complex (8), 60S complex (7), 40S complex (7). In contrast, by mean engaged fraction, the leaders are small synaptic signaling complexes: LRRC7–CAMK2A–ACTN4 (0.313), ELK1–SRF–ELK3 (0.309), PKA–AKAP79–CaN (0.285), PSD95–FYN–NR2A (0.272), BKCa–β2AR–AKAP79 (0.268). Most node-variable by standard deviation (sd) of engaged fraction are: LRRC7–CAMK2A–ACTN4 (0.357), GABA-A receptor (0.344), MR–UBC9–SRC1 (0.327). With 675 of 8,695 proteins reached (7.8%), a complex needs almost all of its subunits inside the reached set to clear BH over 1,132 tests. Two arms with more power were therefore added and are reported alongside it: a set arm over complexes with ≥ half their subunits engaged (median 49 per node), and a continuous arm over the engaged-fraction profile across all 1,132 (Figure 6).

Correlation with the connectome analyses were performed employing the node-by-node similarity matrix design against the 82×82 connectome over the upper triangle (3,321 pairs), Spearman correlation analysis, p from 4,999 joint node-label permutations (seed 20260921). Each arm was also partialled on inter-centroid Euclidean distance, because the connectome is distance-dependent (distance itself: $\rho = -0.209$). Two scopes were analyzed: all 82 nodes, and the 77 whose cascade returns something. An all-empty row is Jaccard 0 against every other node, a similarity value the cascade never measured for 5 nodes. Results of these analyses are summarized in the Figure 6 and Tables 6b and 6c.

**Table 6c.** Correlation of CORUM multiprotein complexes with the connectome analyses summary statistics.

| arm | all82 ρ (p) | all82 partial (p) | live77 ρ (p) | live77 partial (p) |
|---|---|---|---|---|
| **marker panels (reference)** | +0.262 (0.0002) | +0.217 (0.0002) | +0.271 (0.0002) | +0.227 (0.0002) |
| **dimer identities, panel partners** | +0.160 (0.0014) | +0.137 (0.0068) | +0.227 (0.0002) | +0.191 (0.0002) |
| **dimer identities, CNS monomer pool** | +0.120 (0.027) | +0.107 (0.068 ns) | +0.189 (0.0004) | +0.156 (0.0006) |
| **engaged proteins (cascade products)** | +0.157 (0.0028) | +0.136 (0.013) | +0.236 (0.0002) | +0.195 (0.0002) |
| **CORUM, ≥ half subunits engaged** | +0.152 (0.0012) | +0.136 (0.0060) | +0.210 (0.0002) | +0.183 (0.0002) |
| **CORUM, q < 0.05 sets** | −0.001 (0.98 ns) | −0.019 (0.64 ns) | +0.005 (0.91 ns) | −0.016 (0.71 ns) |
| **CORUM engaged-fraction profile** | **+0.265 (0.0002)** | **+0.243 (0.0002)** | **+0.314 (0.0002)** | **+0.287 (0.0002)** |
| **distance (reference)** | −0.209 | — | −0.205 | — |

Correlation of CORUM multiprotein complexes with the connectome are positive and permutation-significant, and statistical significance tests survive the removal of spatial distance (Table 6c). Dimer identity tracks the connectome at +0.14 (82 nodes) to +0.19 (77 nodes) after distance is partialled out. The complex-level profile is the strongest correlation with the brain connectome arm. At +0.243 on 82 nodes and +0.287 on 77, the engaged-fraction profile over all 1,132 CORUM complexes scored slightly above the marker panel it descends from. Thus, compressing 4,000 marker genes into 1,132 complex-occupancy values loses nothing of the connectome relationship and appears to sharpen it.

Of note, all CORUM profiles derive from the marker protein panels and PPI-cascade products as the other axes. Node-specific PPI cascades are set in size and content by the HAP constituents TF-HUB proteins family (H350) of the node's marker panel, in contrast with the HAP constituents of the principal-initiator TF family. The dimer repertoire a brain node can build, and the CORUM complexes it can contain, both track how strongly that node is connected to other nodes. These association remain statistically significant after spatial distance is removed and against a node-label permutation null. Overall, results of these analyses document the contribution of the proteomics vector of the PPI cascade constituents to brain connectome - HAP of EMC associations by demonstrating that node-specific dimer repertoires and CORUM complex occupancy both track the interregional brain connectivity, and the complex-level profile tracks it most closely (Figure 6; Table 6a; b; c).

**Molecular correlates of the mouse brain connectome.**

The fundamental postulate at the foundation of the CSDX model is based on the hypothesis that a human brain region and its named mouse counterpart occupy similar connectivity networks' positions corresponding to an evolutionary conserved architectural design of the mammalian brain connectome. This postulate implies that a projection strength measured between two mouse structures is informative about the connections between the two human structures paired with them. One of the approaches by implementation of which the validity of this hypothesis could be tested is the identification of the molecular correlates of the mouse brain connectome employing gene expression measurements and brain node connectivity data acquired directly from the analyses of the mouse brain.

These analyses revealed that mouse gene-expression identity profiles are positively associated with the mouse projection matrix in all three coverage-qualified parcellations (Table 7; Supplementary Information: Mouse transcriptome–connectome comparison).

**Table 7.** Matched identity tests of mouse and human transcriptome–connectome associations' comparisons. All whole-panel tests have two-sided node-permutation p = 0.0002, including distance-adjusted tests.

| Original map | Strict nodes | Shared genes | Mouse whole-panel ρ | Human whole-panel ρ | Mouse ρ after mouse CCF distance | Human ρ after human MNI distance |
|---|---|---|---|---|---|---|
| **73** | 71 | 8,984 | 0.421 | 0.230 | 0.283 | 0.173 |
| **82** | 81 | 8,978 | 0.373 | 0.241 | 0.212 | 0.191 |
| **92*** | 88 | 8,919 | 0.381 | 0.245 | 0.215 | 0.190 |

Legend: Each species' measured genes were z-scored across the included nodes, and the top 4,000 genes in each node formed its panel. Pair identity was 100 × panel intersection / smaller panel size. Pair identity was compared with the unmodified mouse projection matrix by Spearman Mantel statistic and 4,999 joint node-label permutations; mouse ISH and human AHBA were processed identically on each shared orthologue universe. * 92 nodes sensitivity, without AUDpo. Detailed results of these analyses are reported in the Supplementary Materials.

The conclusion that the same direction of the positive transcriptome/connectome association is present in both species was retested in the matched identity model when human AHBA gene expression data and mouse ISH gene expression data were retrieved and analyzed on the same nodes and the same one-to-one orthologue genes (Table 7). Of note, all whole-panel tests in the Table 7, including distance-adjusted tests, have two-sided node-permutation p = 0.0002. These findings are consistent with the conclusion of evolutionary conservation of a broad transcriptomic-identity/connectivity association.

A matched 5-predictor, rank-pair multiple-regression/commonality analysis was run using panel-gene breadth, low expression, anterior–posterior gradient, HAP membership and within-species centroid distance, with 4,999 node permutations. Full-model $R^2$ for mouse/human relation is 0.220/0.153 (70-node), 0.200/0.149 (80-node), and 0.209/0.165 (87-node 92 sensitivity). The mouse AP-gradient predictor has unique $R^2$ = 0.0095/0.0104/0.0129 (uncorrected permutation p = 0.011/0.005/0.001), while mouse HAP membership has unique $R^2$ = 0.0090/0.0054/0.0039 (p = 0.056/0.113/0.153). The 73/82/88 mouse HAP-family identities are usually positively related to projections before conditioning. However, controlling for whole-panel identity changes the direction for some families; the same control is often negative in the matched human arm. Mouse C171 family has partial-panel empirical p = 0.015 / 0.005 / 0.025 at 71/81/88 nodes, respectively; while mouse H350 family has p = 0.015 / 0.015 / 0.010 (detailed results of these analyses are reported in the Supplementary Materials).

Overall, results of these analyses provide the initial support of a broad evolutionary-conserved transcriptomic-identity/brain connectivity association: mouse expression similarity tracks the mouse brain projections, and the matched AHBA human transcriptome identity/brain connectome association has the same direction.

**Analyses of matched seven-axis mouse gene expression and human gene expression models against the brain connectome.**

Human AHBA gene expression-derived and mouse ISH gene expression-derived seven-predictor models were fitted against the same unchanged mouse projection outcomes, on identical anatomical nodes and identical one-to-one orthologue gene universes within each of the 73-, 82- and 92-node CSDX maps (Table 8; see Methods and Supplementary Information for detailed descriptions of the comparative analyses of human-mouse data sets).

**Table 8.** Matched seven-axis mouse and human gene expression models against the brain connectome.

| Source map → tested nodes | Ortholog genes | Expression arm → mouse outcome | Five-axis $R^2$ | Seven-axis $R^2$ | Five→seven $\Delta R^2$ | Held-node seven-axis $R^2$ |
|---|---|---|---|---|---|---|
| **73 → 70** | 8,984 | Mouse ISH | .2178 | .2321 | .0143 | .1947 |
| **73 → 70** | 8,984 | Human AHBA | .1499 | .1643 | .0145 | .1201 |
| **82 → 80** | 8,978 | Mouse ISH | .1995 | .2099 | .0104 | .1783 |
| **82 → 80** | 8,978 | Human AHBA | .1476 | .1725 | .0249 | .1382 |
| **92 → 87** | 8,919 | Mouse ISH | .2086 | .2235 | .0148 | .1956 |
| **92 → 87** | 8,919 | Human AHBA | .1642 | .1873 | .0232 | .1552 |

Legend: Unique $R^2$ is the full seven-predictor $R^2$ minus $R^2$ without that axis. One-sided empirical p uses 4,999 joint node-label permutations of the mouse-projection outcome. The figure's q is BH over 42 axis tests (7 predictors × 3 maps × 2 expression species) for the continuous-CORUM model. The binary-CORUM sensitivity has its own 42-test family; q over 21 tests within each expression arm is additionally available in the axis table to connect with the attached figure's original correction. Each arm was also assessed by pooled leave-one-node-out prediction of ranked node pairs.

Predictors were designated as follows: B = scaled breadth; A = low abundance; R = anterior–posterior gradient; H = four HAP family union; D = centroid distance; I = candidate-dimer repertoire; C = CORUM profile. In the primary matched comparison, the human expression arm has larger unique anterior–posterior-gradient $R^2$ on all three maps and a positive HAP-family unique term in the 82/92 nodes maps after the BH correction. In the per-predictor unique terms analysis, the unique $R^2$ value was estimated for each entry (node-permutation p; BH q across 42 tests). The human R (anterior-posterior gradient) bar clears the joint 42-test correction on each map; human H q value (HAP family constituents) clears it for the 82 and 92 maps, while the 73 H q is marginal .0552. Human I bar (reflecting the value for candidate dimer repertoire) has nominal p = .0168/.0178 in the latter two maps, but q = .0748, so the adjusted p value result does not formally establish an individual human dimer-axis contribution in a matched model. In the ≥50% CORUM-set sensitivity, neither species' dimer axis clears $q < .05$.

Employing the most conservative 82-node CSDX setting (Methods), the decomposition analyses were carried out to calculate the fractions of the unique and shared contributions of the seven-predictor models to evaluate their within-model fractions (Figure 7; Table 9). Both human and mouser models use the same 80 mouse-projection nodes. The mouse ISH gene expression model has 82.9% net shared and 17.1% unique fitted $R^2$, while the human AHBA gene expression model has 45.6% net shared and 54.4% unique fitted $R^2$ (Figure 7; Table 9). The four-panel figure (Figure 7) shows seven bars for each model, followed by each model's shared/unique partition. Of note, the mouse arm uses 10,567 genes and the human arm uses 8,978 genes; therefore, their fractions are within-model descriptions of decomposition analyses results, not a matched species comparison.

**Table 9.** Independent decomposition analyses of seven-axis mouse and human gene expression models against the brain connectome employing the most conservative 82 nodes CSDX setting.

| 82-node source model→ 80 analyzed nodes | Fitted full $R^2$ | Unique sum $R^2$ | Unique fraction | Net shared $R^2$ | Net shared fraction | Positive / negative 2–7-way components |
|---|---|---|---|---|---|---|
| **Human AHBA → mouse projections, 8,978 genes** | .172485 | .093756 | **54.36%** | .078729 | **45.64%** | +.200739 / −.122010 |
| **Mouse ISH → mouse projections, 10,567 genes** | .219447 | .037440 | **17.06%** | .182008 | **82.94%** | +.205198 / −.023190 |

| Axis | Human unique $R^2$ | Human fraction of its full $R^2$ | Mouse unique $R^2$ | Mouse fraction of its full $R^2$ |
|---|---|---|---|---|
| **B: breadth** | .003389 | 1.96% | .000350 | 0.16% |
| **A: low abundance** | .000008 | 0.00% | .001365 | 0.62% |
| **R: AP gradient** | .028346 | 16.43% | .004891 | 2.23% |
| **H: four-family membership** | .021079 | 12.22% | .000572 | 0.26% |
| **D: centroid distance** | .014157 | 8.21% | .009413 | 4.29% |
| **I: candidate dimers** | .024270 | 14.07% | .020764 | 9.46% |
| **C: CORUM profile** | .002507 | 1.45% | .000085 | 0.04% |

Legend: The analyses use the mouse-ISH reference analysis (10,567 orthologues) and the human AHBA 82-map seven-axis analysis (8,978 joint mouse–human orthologues). Both evaluate the same 80-node mouse projection outcome, but the mouse and human molecular gene universes and candidate cascades are different, indicating that model fractions are within-model descriptions, not a matched effect-size contrast or a species-difference test.

For each predictor, the unique $R^2$ value was estimated as the value of fitted seven-axis $R^2$ subtracting fitted six-axis $R^2$ after removing that predictor. Its fraction is calculated as percentile of unique $R^2$ / full seven-axis $R^2$. The human AP gradient (p=.0002, q=.0014), HAP membership (p=.0040, q=.0185) and candidate dimer repertoire (p=.0168, q=.0415) all meet within-arm BH correction over the 21 axes × maps tests; the mouse candidate-dimer term has a nominal p = .016 but q=.120 and none of its seven unique terms meets BH q<.05. The paired analysis of formal between-species differences found no statistically-supported species-arm difference for any of the seven predictors across the three matched parcellations (Supplementary Information). The largest descriptive differences favor the human AHBA arm for the AP-gradient (designated R), HAP-family union (designated H) and candidate-dimer (designated I) axes, but their node-level uncertainty level includes zero. Therefore, larger predictive values of human AP-gradient, PPI cascade HAP family constituents, and candidate-dimer should be regarded as descriptive between-species differences, not statistically-supported established between-arm effects.

Differences of the aggregate unique and net shared fractions were confirmed employing three exactly matched between species models in which both human and mouse arms use the exactly same selections of nodes/genes/projection data: 70; 80; 87 nodes; 8,984; 8,978; 8,919 one-to-one orthologues; and corresponding original mouse projection data (Table 10).

**Table 10.** Analysis of the aggregate unique and net shared fractions employing exactly matched between species three parcellations' models.

| Source → nodes | Mouse unique / net shared | Human unique / net shared | Human − mouse unique fraction | Basic 95% interval | Paired p | BH q (n=9) |
|---|---|---|---|---|---|---|
| **73 → 70** | 7.95% / 92.05% | 48.87% / 51.13% | +40.92 PP | +18.43 to +66.32 | .002 | **.018** |
| **82 → 80** | 10.08% / 89.92% | 54.36% / 45.64% | +44.28 PP | +13.06 to +78.25 | .010 | **.030** |
| **92 → 87** | 10.86% / 89.14% | 48.75% / 51.25% | +37.89 PP | +11.29 to +67.62 | .008 | **.030** |

Legend: Each arm's unique sum is the sum of its seven full-minus-drop-one $R^2$ values; net shared $R^2$ = full fitted $R^2$ − unique sum. The unique fraction is unique sum/full $R^2$, and its net shared fraction is exactly 1 − unique fraction. PP, percentage points.

Results reported in the Table 10 demonstrate that in all three parcellations of the CSDX model the differences between the unique and net shared fractions pass BH $q < 0.05$, suggesting that these between species descriptive differences are statistically significant. The absolute values of unique fraction sums (reported as the mouse → human values of $R^2$ units) are: 0.0185→0.0803; 0.0211→0.0938; and 0.0243→0.0913 (in the CSDX models of 70; 80; 87 nodes, respectfully). The corresponding human-minus-mouse differences are: +0.0619; +0.0726; +0.0671. The absolute values of net shared mouse → human balances are: 0.2137→0.0840; 0.1887→0.0787; and 0.1992→0.0960 (their corresponding nominal paired p are: 0.001; 0.001; 0.002). Results of the independent sensitivity analysis with contributions of ≥50%-engaged binary-CORUM complexes (Figure 8) show significant human-mouse fraction differences of +42.7; +46.4; +41.1 percentage points (corresponding BH q values are 0.009; 0.030; 0.018).

Overall, these observations are compatible with the hypothesis of a conserved brain regions-defined transcriptome–connectome associations. Comparisons of the seven-arm human and mouse transcriptome/connectome association models demonstrate that the mouse-derived model generally fits the mouse brain connectome better, while the human-derived model assigns a significantly larger fraction of its transcriptome/connectome association to unique rather than net shared contributions.

**Discussion.**

In this contribution, the cross-species digital xenocortication (CSDX) model based on a pairing of a human brain structure with one or more mouse brain structures was conceived, constructed, and implemented to carry out the analytical experiments aimed at dissection of the molecular correlates of the human brain connectome. A major focus of this study was centered on the systematic exploration of potential regulatory associations of the four families of PPI constituents of Hybrid Assembly Pathways (HAP) of endogenous multiprotein complexes (EMC) in relation to between-nodes connectivity patterns of 92 anatomically distinct human brain regions. In the main arm of this study, gene expression for the analyzed nodes was measured in the human brain, while connection strength between two nodes was measured in the mouse brain. In the second arm of this investigation, both gene expression and connection strength were measured in the mouse brain.
Mouse ISH-defined gene-expression identity is positively associated with the mouse brain nodes' projection matrix in all three coverage-qualified parcellations. The same direction is present when AHBA expression is retested on the same nodes and the same matched one-to-one orthologue genes. These observations support the hypothesis of conservation of a broad transcriptomic-identity/connectivity association in mammalian brains.

Previous studies reported analytical evidence supporting the concept of global, chromosome-naïve genomic regulatory networks that are engaging in their operation 1,122 constituents of HAP of EMC (Introduction). It has been suggested that this multi-vectored multi-component biological system was designed to facilitate the maintenance of key elements of the cellular homeostasis with the principal contribution to primates' brain development, including the apparently human-specific regulatory impacts on development of ~200 anatomically-distinct brain regions (Glinsky, 2022-2026a; b; c). However, the potential roles of HAP constituents in building the structural-functional elements of adult human brain remain largely unknown. In this contribution, the associations with the brain connectome of transcriptomics and proteomics elements acting via distinct molecular mechanisms attributed to the PPI cascade constituents were examined. Specifically, potential contributions of two principally distinct mechanisms of the PPI constituents' activities that are conditioned on the chromatin states of the genomic regions in which specific sets of TE-encoded DNA loci are located were evaluated.

TE-encoded DNA loci located in the transcriptionally-silent heterochromatin regions function as DNA templates for initiation of the PPI cascade chains creating the cell type-specific intracellular pools of protein dimers, thus nucleating the protein dimerization reactions for EMC assembly (Glinsky, 2026a; b; c). This transcriptionally-silent heterochromatin-associated vector of the activity of PPI constituents was further dissected through analyses of 2 distinct modes of HAP operations with biologically distinct endpoints, namely TF-TF PPI cascade branch and TF-HUB proteins PPI cascade branch, which appear to manifest distinct association patterns with the brain connectome. These 2 distinct operational modes could be viewed as congruent secondary vectors of biological functions of the PPI cascade constituents bound to the DNA loci located within transcriptionally silent heterochromatin.

In contrast, in the contexts of transcriptionally active accessible chromatin regions, TE-encoded DNA loci exert their regulatory functions by acting as promoters, enhancers, non-coding RNA molecules, and/or boundary elements of topologically associating domains of 3D structures of the linear chromatin fibers and execute a wide-range of regulatory functions via expression changes of specific sets of target genes (Lu et al., 2014; Elbarbary et al., 2016; Sundaram and Wysocka, 2020; Rosspopoff et al., 2025; Glinsky, 2015 - 2026). Thus, regulatory actions of the PPI cascade constituents bound to DNA loci located in transcriptionally active genomic regions could be viewed through the lenses of multiple mechanistically-distinct secondary transcriptomics vectors of their biological functions, which were analyzed in this contribution via the assessments of the association of down-stream target genes of PPI cascade constituents with the brain connectome.

Per-node availability analyses revealed the common key features of the molecular architecture of the four families of PPI cascade constituents within different contexts of intracellular proteomes of three CSDX model parcellations: a) ubiquitous expression; b) balanced representation; c) universal availability and engagements (Figures 1 and 2). These results recapitulate the findings of the investigation of the constituents of PPI cascades carried out within the context of cell type-specific intracellular protein monomer pools documented in the Human Protein Atlas (HPA) of 154 human cell types (Karlsson et al., 2021; Glinsky, 2026).

Analyses of three parcellations of the CSDX model revealed highly reproducible profiles of absolute Spearman Mantel correlations between node-pair detection-pattern similarity and mouse connectivity (Figure 2; A4, B4, C4 panels; asterisks denote permutation $p < 0.05$. (A5, B5, C5) Testing whether the anatomical placement of detection carries signal beyond detection breadth demonstrated that observed correlations (compared with a breadth-matched null that replaces each family member with a non-member detected in the same number of nodes) remain statistically significant for constituents of the principal initiator TF family (P545); initiator TFs; and the union of PPI cascade constituents (Figure 2; A5; B5; C5 panels). Analyses of HAP constituents identity and cascade engagement against the between-nodes' projection strength documented statistically significant associations (Figure 3C; asterisks denote $p < 0.05$, 4,999 permutations).

Delineations of transcriptional correlates of the human brain connectome are in alignment with the hypothesis that the PPI cascade HAP constituents that are expressed in a cell constitute integral components of the intracellular transcriptome and proteome. It has been observed that transcriptional identity tracks measured projection strength across the 2,628 pairs at $\rho = +0.261$, $p = 0.0002$, and it survives the spatial comparison control: partialling inter-centroid distance leaves +0.202 at the same p value (Figure 4; panel B1). Thus, brain areas that project more strongly to each other look alike transcriptionally, and not only because they are near each other. Several separately defined transcriptome axes each beat their rank-matched controls and no single transcriptome subset reproduces the full transcriptome panel. A combination of redundant and distributed contributing axes is exactly recapitulate features of what a multifactorial identity association signal looks like. Therefore, the relation between the transcriptome and connectome is multifactorial because no single gene expression property accounts for it in full and several separately defined predictors each recover the statistically-significant relation. In agreement with this conclusion, it has been observed that transcriptional identity predictor computed from the genetic targets of the PPI cascade HAP families tracks the brain regions projection strength as strongly as the whole-transcriptome identity ($\rho = +0.25$ to $+0.28$; $p = 0.0002$, distance-independent), but appears to display no significant association signal once whole-panel identity is partialled out.

Analysis of the relations of node-specific dimer repertoires and CORUM complex occupancy and brain connectivity document the contribution of the proteomics vector of the activity of PPI cascade constituents to the associations with the brain connectome by demonstrating that node-specific dimer repertoires and CORUM complex occupancy both track the interregional brain connectivity, and the complex-level profile tracks it most closely (Figure 6; Tables 6a; b; c). The brain connectome defined by the measurements of how strongly a node is connected to other nodes is strongly associated with protein dimer repertoire a brain node can build and the CORUM complexes it can contain. Both of these PPI cascade constituents-defined proteomics vectors track the brain connectivity patterns reflecting how strongly anatomically-defined brain nodes are connected to each other. Importantly, these associations remain statistically significant after spatial distance is removed and against a node-label permutation null.

**Conclusions.**

Identification and characterization of the molecular correlates of the human brain connectome continue to represent a highly significant challenge due to ethical considerations, technical constraints, and analytical limitations. In this contribution, to address some of these problems the cross-species digital xenocortication (CSDX) model based on a pairing of a human brain structure with one or more mouse brain structures has been conceived, designed, and implemented. The reported CSDX model tested the hypothesis that a human brain region and its named mouse counterpart occupy corresponding positions in an evolutionary conserved mammalian brain connectome plan, so that a projection strength measured between two mouse structures is informative about the relation between the two human structures paired with them. In one arm of the CSDX model, gene expression for the node was measured in the human brain, while connection strength between two nodes was measured in the mouse brain. In another arm of the CSDX model, both gene expression for the node and the connection strength between nodes were measured in the mouse brain. Consistently, both arms of the CSDX model documented strong positive associations of brain connectome patterns with node-specific transcriptional profiles independently defined for each species. However, the mouse-only model appears to fit the mouse brain-measured connectome better, while the human brain nodes gene expression-derived model assigns a significantly larger fraction of its transcriptome/connectome association to unique rather than shared contributions. In the human arm of the CSDX model, the gene expression gradients (defined as rostrocaudal or anterior-posterior axes) and HAP constituency were identified as most significant unique contributors of the multifactorial model tracking the associations with the brain connectome. Overall, observations reported in this contribution provide strong support for the hypothesis of conservation of a broad transcriptomic-identity/connectivity association in mammalian brains. Employing the CSDX model, the statistically significant associations with the brain connectome of transcriptomics and proteomics regulatory elements acting via distinct molecular mechanisms attributed to the four families of PPI cascade constituents were documented. Specifically, potential contributions of two principally distinct mechanisms of the PPI constituents' activities that are conditioned on the chromatin states of the genomic regions in which specific sets of TE-encoded DNA loci are located were demonstrated. Both analytical arms of developed and validated in this contribution the CSDX model of the mammalian brain connectome are freely available for the scientific community. Present analyses nominate elite panels of top-scoring candidate protein dimers and CORUM multiprotein complexes for follow-up mechanistic studies of their potential roles in structural-functional features governing mammalian brain connectome.

## Methods

### Cross-species brain-area model and expression panels

Human brain-area gene expression was compared with projection strength measured in the mouse brain through manually curated human–mouse anatomical correspondences. The analysis assumed that paired regions occupied corresponding positions in a conserved connectional plan; it did not measure human projection strength. A 73-node base map included tier A/B correspondences supported by at least 12 Allen Human Brain Atlas (AHBA) samples from four of six donors. A finer, 92-node map subdivided eligible regions, and an 82-node map retained its A/B-graded correspondences. Node construction, homology grading, projection pooling and symmetrization are described in the Supplementary Methods.

Within each donor, one probe per gene was selected by mean intensity, expression was standardized by gene across samples, and standardized values were averaged within nodes and then across contributing donors. The resulting matrix contained 20,737 genes. For each node, genes were ranked by expression relative to that gene's distribution across nodes; the top 4,000 formed its marker panel. As a separate availability measure, a gene was called detected in a node when its AHBA present call occurred in at least 50% of that node's pooled samples. Analyses compared marker-panel membership, detection, and their intersection without treating these rules as interchangeable.

### Protein-network products and complex availability

Author-defined sets comprised 545 principal and 171 complementary initiators, 245 TF–TF targets, 350 hub proteins, and their 1,122-protein union. Available initiators seeded separate cascades through the Enrichr Transcription_Factor_PPIs and PPI_Hub_Proteins libraries. At each iteration, one-sided hypergeometric enrichment was tested against the branch-specific available background, with Benjamini–Hochberg correction at $q < 0.05$; significant terms seeded the next iteration until convergence or 12 levels. Reached proteins were distinguished from proteins merely present in a node's marker panel. Binary reached-member profiles were compared across nodes and clustered to assess recovery of anatomical divisions.

A separate direct-partner analysis selected HAP-family seeds from each node's marker panel and paired them with documented PPI partners available in a central-nervous-system (CNS)-expressed background. This background contained 8,695 genes expressed at nTPM $\geq 1$ in at least one of eight Human Protein Atlas CNS cell types. Matched controls drew non-HAP seeds from the same node panel while matching documented partner count. For the 82-node analysis, dimers were enumerated both with partners restricted to the node's panel and with partners drawn from the CNS-expressed background; a dimer was classed as engaged when both partners were cascade products. CORUM complexes with at least three subunits were evaluated by engaged-subunit fraction and by hypergeometric enrichment against the CNS-expressed background, with Benjamini–Hochberg correction across tested complexes.

**Connectome comparisons and controls**

Node-by-node similarities of marker panels, family-restricted panels, reached proteins, dimers and complex profiles were compared with symmetrized mouse normalized projection volume over unordered node pairs using Spearman Mantel statistics. At 73 nodes, family-restricted identity was overlap normalized by the smaller panel; the finer-parcellation analyses used Jaccard similarity where specified. Partial analyses rank-residualized similarities and projection strengths on inter-centroid distance or whole-marker-panel identity. Significance was estimated by jointly permuting node labels on the rows and columns of a similarity matrix (4,999 permutations). Anatomical-division recovery was measured by adjusted Rand index against 2,000 division-label permutations.

To test specificity, family identity was compared with size-matched non-HAP gene sets. Detection-pattern tests additionally used gene-by-gene controls matched on the number of nodes detecting each member. Regulatory-target identity was constructed from JASPAR 2025 motif and ChEA 2022 ChIP target sets; subtraction analyses removed HAP constituents, documented targets, or both from marker panels before repeating the connectome comparison. Five-predictor commonality analysis partitioned rank-based connectome variance among gene-subset similarities defined by detection breadth, abundance, rostrocaudal gradient and HAP membership, and inter-centroid distance. Unique components were tested by node-label permutation. Finally, division-separation analyses were repeated after restricting the 82- and 92-node maps to precisely remapped nodes, using the same breadth-matched null while recording the change in division composition.

**Mouse transcriptome–connectome comparison: coverage-qualified analysis**

**Data and harmonization**

- The original 73/82/92 node maps and mouse normalized-projection-volume matrices were left unchanged (Supplementary Methods). For the mouse expression arm, Fulcher's Allen mouse ISH matrix supplies 19,419 genes × 213 structures; Allen StructureUnionize API records, restricted to experiments listed in the October 2014 mouse-expression index, supply missing some mapped structures. Assay-level expression energy was averaged within gene and structure, then equally across the structures assigned to each node (ignoring missing gene–structure assays). The two sources agreed exactly for 99.73% and 99.79% of ~19,000 overlapping gene–structure observations in two tested structures (Pearson 0.99980 and 0.99985); individual discrepancies remain. These are ISH energy, not RNA-seq abundance or AHBA microarray detection calls.
- Mouse–human matching uses one-to-one mouse/human orthologue groups from MGI's Mouse/Human Homology Report, linked by mouse Entrez ID to the Allen gene ID; neither capitalization nor name resemblance was treated as orthologues. Matching further requires a gene measured at every included node in both species. The resulting shared universes contain 8,984 genes at 71/73 nodes, 8,978 at 81/82 nodes, and 4,838 at 89/92 nodes. Omitting the assay-poor AUDpo node from the last analysis increases coverage to 8,919 genes at 88/92 nodes.

- Nodes with every listed mouse substructure covered by the archived ISH data: 71/73, 81/82, 89/92. The 73-node extrastriate visual node lacks VISli and VISpor; the oculomotor/trochlear node lacks IV in every resolution. In the 92 map, VISli is entirely unmeasured and the VISpor-containing node is incomplete. AUDpo has only 7,654 of 19,419 MAT genes with archive-matched ISH data. The expression_coverage_qc.csv table (Supplementary Information) gives gene-level completeness and the precise exclusions. Retaining partially sampled substructures as if their original node definitions were complete would overstate coverage.

**Design and implementations of matched identity tests**

Each species' measured genes were z-scored across the included nodes, and the top 4,000 genes in each node formed its panel. Pair identity was 100 × panel intersection / smaller panel size. Pair identity was compared with the unmodified mouse projection matrix by Spearman Mantel statistic and 4,999 joint node-label permutations; mouse ISH and human AHBA were processed identically on each shared orthologue universe (Table 8; refer to the Supplementary Information for detailed descriptions of the comparisons of human-mouse data sets). This matched rerun is not numerically interchangeable with the CSDX model full-node, full-gene AHBA run.

All whole-panel tests in this table have two-sided node-permutation $p = 0.0002$, including distance-adjusted tests. Mouse distances use CCF reference space 9 centers; human distances use folded-hemisphere AHBA MNI centroids. The mapped ENTmv has no supplied CCF center: distance-adjusted tests omit its entire node (70/80/88/87 nodes, respectively), whereas unadjusted tests use the strict node counts shown. Thus raw and adjusted $\rho$ values are not evaluated on exactly the same nodes.

**HAP-family specificity and variance**

- The 73/82/88 mouse HAP-family identities are usually positively related to projections before conditioning. However, controlling for whole-panel identity changes the direction for some families; the same control is often negative in the matched human arm. Broad identity conservation does not entail family-specific conservation. The 200-draw size-matched random-set controls are reported in matched_random_set_control.csv table (Supplementary Information): mouse C171 has partial-panel empirical $p = 0.015 / 0.005 / 0.025$ at 71/81/88 nodes; mouse H350 has 0.015 / 0.015 / 0.010. These are uncorrected exploratory values over six overlapping family tests and multiple resolutions.
- A matched 5-predictor, rank-pair multiple-regression/commonality analysis was run using panel-gene breadth, low expression, anterior–posterior gradient, HAP membership and within-species centroid distance, with 4,999 node permutations. Full-model $R^2$ for mouse/human is 0.220/0.153 (70-node), 0.200/0.149 (80-node), and 0.209/0.165 (87-node 92 sensitivity). The mouse AP-gradient predictor has unique $R^2 = 0.0095/0.0104/0.0129$ (uncorrected permutation $p = 0.011/0.005/0.001$), while mouse HAP membership has unique $R^2 = 0.0090/0.0054/0.0039$ ($p = 0.056/0.113/0.153$). These pairwise $R^2$ values are descriptive under node permutation, not independent pair observations, and the differences between species were not formally tested.

**Scope, limitations, and next requirements**

Analyses of these deliverables were designed to test the reproducibility of identity, distance, random-set and variance-decomposition components of the AHBA protocol on a matched mouse-expression arm. It does not tested the full scope of independent reproducibility of the AHBA per-sample detection, cascade engagement, CORUM-complex tests or the full exact node counts. ISH expression energy cannot be treated as the AHBA sample-detection fraction, and mouse protein/cascade membership would require additional experiments to validate orthologization protocol and a separately justified detection rule. VISli and VISpor have **zero** unionizes at their mapped structure IDs in the queried Allen API, while IV has no experiments in the 2014 index; full-node analyses therefore require a different, spatially resolved expression source at those exact territories, not interpolation or silent parent substitution. Analyses of mouse–human differences mix species, assay platform, structure definitions and the shared mouse projection outcome; no fully validated claims of species-specific effect magnitude follow from the two ρ values alone.

**Sources.** Fulcher, Allen Brain Atlas mouse gene-expression matrix; Allen 2014 mouse-expression experiment index; Allen RMA API; Allen CCF9 structure centers; MGI mouse/human homology report. The unmodified mapping/projection provenance is documented separately in the Supplementary Information: supplementary_methods_parcellations.md.

**Acknowledgements**

This is the single author contribution. All analyses were conceived, prompted and initiated by the author. Implementation and execution of the analyses were carried out with the assistance of Academia Co-scientist (Anthropic Claude, model Opus 5; GPT-6 Sol), operating on the primary data files under the author's specification; all reported values were verified by the author against those files. Academia Co-scientist (Anthropic Claude, model Opus 5; GPT-6 Sol) was used for computational analysis, figure generation and initial drafting of parts of the Introduction and Methods. All outputs were specified, verified and edited by the author, who takes full responsibility for the manuscript. The AI tool is not listed as an author and made no authorship-level contribution.

The authors used Academia Co-scientist (GPT-6 Sol; accessed October 2026) to assist with integrating public protein-annotation datasets, writing analysis code, and preparing preliminary draft figures and captions. The author used different AI assistants (Academia Co-scientist GPT-6 Sol and Anthropic Claude, model Opus 5) for independent cross-validation of the computational analyses outputs generated by the AI assistants. The reproducibility of the AI assistants' outputs starting the task of complex multi-step computational analyses of 154 human cell types from the independent retrievals of the public database records exceeds 99.5%. The authors retain responsibility for verifying the data, analyses, references, and final interpretation.

**Conflict of Interest**

The author declares that he has no conflicts of interest with the contents of this article.

**Figure legends.**

**Figure 1. AHBA-supported brain-area nodes and per-node availability of HAP protein families. (A1)** Numbers of Allen Human Brain Atlas (AHBA) samples pooled into each of 109 proposed tier A or B nodes, ordered by sample count and coloured by anatomical division. The dashed line marks the 12-sample threshold. Seventy-three nodes meet both this threshold and the requirement for samples from at least four of six donors; 74 meet the sample threshold alone. **(A2)** Numbers of connectome nodes and unordered node pairs under the existing mouse-matched lobe panels, a simulated mouse18 lobe mapping, the approved AAL3 human set, and two area-level sample thresholds. The 73-node analysis contains 2,628 pairs. **(A3)** Anatomical-division composition of those 73 nodes. **(B1–B9)** Per-node counts for the P545, C171, T245 and H350 families, the 716-initiator set, and their 1,122-protein union under three availability rules (rows) and 73-, 82- and 92-node parcellations (columns). The top row uses each node's top-4,000 marker panel without a detection filter. The middle row counts genes with a present AHBA detection call in at least 50% of that node's samples, pooled across contributing donors. The bottom row requires both marker-panel membership and detection. Solid lines give observed family-member counts; dotted lines give the expectation for each family from the same drawn gene set. All rules use the same population of 20,737 attempted genes. Nodes are ordered separately in each plot by availability of the 1,122-protein union, so horizontal positions cannot be compared across plots. The drawn set is fixed at 4,000 genes in the top row but varies by node under the detection-based rules.

**Figure 2. Detection-based availability of HAP protein families across three brain-area parcellations.** Panels **A**, **B** and **C** repeat the same analyses for 73-, 82- and 92-node connectomes, respectively. A gene is detected in a node when its AHBA present call occurs in at least 50% of that node's samples, pooled across contributing donors. Unlike the node-specific top-4,000 marker-panel rule, this absolute threshold permits detection in every node. All tests use the same population of 20,737 genes attempted on the array. **(A1, B1, C1)** Distributions of the number of nodes detecting each measured family member, compared with all attempted genes; the dashed line marks detection in every node. **(A2, B2, C2)** Observed-to-expected family-member availability across nodes under the top-4,000 marker-panel rule, the ≥50% detection rule, and their intersection. The expectation uses each node's drawn-set size. **(A3, B3, C3)** Numbers of nodes with family enrichment at Benjamini–Hochberg $q < 0.05$ for each rule. **(A4, B4, C4)** Absolute Spearman Mantel correlations between node-pair detection-pattern similarity and mouse connectivity, shown without adjustment and after separately partialling inter-node distance or whole-marker-panel similarity. Asterisks denote permutation $p < 0.05$. **(A5, B5, C5)** Observed correlations compared with a breadth-matched null that replaces each family member with a non-member detected in the same number of nodes, testing whether the anatomical placement of detection carries signal beyond detection breadth. Bars show observed correlations; markers and error bars show the null mean and standard deviation. **(A6, B6, C6)** Family-member count steadiness across nodes and separation between anatomical divisions, expressed as z scores against the same breadth-matched null. Higher buffering scores indicate less across-node variation in member count; positive division-separation scores indicate greater separation than matched genes. The dotted reference line marks $z = 1.96$.

**Figure 3. HAP marker-panel content, cascade reach, and projection connectivity across 73 AHBA area-level nodes.** (**a**) Pale bars show the number of distinct HAP proteins among each node's top-4,000 marker genes, selected by expression ranked relative to that gene's mean across nodes. Coloured bars show the number of HAP proteins reached by the PPI cascade after term enrichment at Benjamini–Hochberg (BH) $q < 0.05$; colours identify anatomical divisions. Numbers above the coloured bars give the reached counts. An asterisk identifies each of seven nodes for which no PPI term passed the enrichment threshold. These nodes contain 121–192 HAP proteins in their marker panels, but their significance-defined cascade reach is zero; marker-panel membership must not be interpreted as cascade reach. (**b**) Adjusted Rand index compares anatomical division with a ten-group clustering of node-level binary reached-member profiles for P545 principal initiators, C171 complementary initiators, T245 TF–TF targets, H350 hub proteins, and their 1,122-protein union. Labels above bars give z scores against 2,000 division-label permutations. (**c**) Spearman Mantel

correlations compare mouse projection strength with HAP-family marker-panel identity or cascade engagement across the same nodes. Light-blue bars show identity correlations after partialling inter-centroid distance; asterisks denote two-sided node-label permutation $p < 0.05$ (4,999 permutations). (**d**) Symmetrized mouse normalized projection volume for the 73 nodes, ordered by anatomical division. Colour represents log10(projection volume + 0.001); white rules separate divisions.

**Figure 4. HAP-restricted transcriptional identity and controls for its association with brain connectivity.** Analyses use 73 AHBA brain-area nodes and 2,628 unordered node pairs. Node-pair transcriptional identity is the overlap of the relevant genes in each node's top-4,000 marker panel, normalized by the smaller panel. (**A1–A4**) Mouse projection strength plotted against identity defined by the 1,122-protein HAP union, P545 principal initiators, T245 TF–TF targets or H350 hub proteins. Each plot uses the same node pairs; annotations report Spearman Mantel correlations before adjustment and after separately partialling inter-centroid distance or whole-marker-panel identity. (**B1–B3**) Whole-marker-panel identity correlates with connectivity, but HAP-family identity closely tracks whole-panel identity and adds little after it is partialled out. Family-restricted correlations are also compared with 2,000 size-matched draws of non-HAP genes; the plotted families do not exceed this null. (**B4**) In contrast to the pairwise identity test, binary profiles of proteins reached by the cascade recover the ten anatomical divisions, measured by adjusted Rand index against 2,000 division-label permutations. (**C1–C2**) Identity defined by ChIP-documented regulatory targets of each HAP family is correlated with connectivity using JASPAR 2025 motifs or ChEA 2022 ChIP targets; little remains after whole-marker-panel identity is partialled out. (**C3–C4**) Independent transcription-factor draws increasingly share the same 2,000-gene target panel as draw size grows, limiting the resolution of a family-specific target comparison. The profiled transcription-factor universe is itself predominantly HAP-associated, leaving a limited non-HAP pool for a matched null. (**D1–D2**) Whole-panel identity remains associated with connectivity after removing HAP constituents, documented HAP-factor targets, or both, shown before and after partialling inter-centroid distance. All plotted conditions have node-label permutation $p = 0.0002$ (4,999 permutations); none of the subtraction conditions lies one standard deviation from its matched random-removal null. Keep-only conditions are shown without a matched null.

**Figure 5. Commonality decomposition of gene-defined connectome associations across three brain-area parcellations.** Multiple regression on rank-transformed node-pair matrices partitions the association with mouse projection strength among five predictors: similarity of marker-panel genes in the intermediate breadth band (B), lowest abundance quartile (A), steepest rostrocaudal-gradient quartile (R), and HAP-constituent set (H), plus inter-centroid distance (D). (A–C) Results for the published 73-area set, 82 A/B-homology nodes and 92-node parcellation, respectively. Breadth bands are 21–40, 24–45 and 26–50 nodes, scaled to the parcellation size. The full-model $R^2$ is 0.1672, 0.1508 and 0.1793, respectively. Left plots show each predictor's individually explained $R^2$, divided into its unique contribution (dark) and the contribution shared with other predictors (light). Percentages and p values beside the bars refer to the unique contribution as a share of full-model $R^2$, tested by 4,999 joint node-label permutations. Unique components account for 41%, 29% and 19% of explained variance in A, B and C; the remainder is shared among predictors. Right plots show commonality coefficients for predictor combinations whose absolute contribution is at least 2% of full-model $R^2$. Positive coefficients indicate shared explanatory variance; negative coefficients indicate suppression, in which predictors jointly explain more than the sum of their separate contributions. Thus, the unique contribution of a gene-defined axis should be distinguished from the larger variance it explains together with other axes.

**Figure 6. Remap precision and node-specific PPI cascade products in the AHBA connectome analysis.** **(A1–A2)** Division-separation z scores against the same breadth-matched, 200-draw null are shown for six HAP-related protein sets. A1 compares all 82 A/B-homology nodes, 53 high-confidence nodes and 61 precisely remapped nodes; A2 compares 29 imprecisely remapped nodes, all 92 nodes and 63 precisely remapped nodes. Precise nodes map an AHBA structure group to one mouse target without splitting; imprecise nodes were formed by dividing a parent group. High confidence is the 82-node map's separate A/B-grade criterion, so the A1 restrictions are not nested. Filled markers denote permutation $p < 0.05$. Restriction to precise nodes raises the P545 principal-initiator statistic from +3.03 to +6.10 in the 82-node map and from +2.13 to +5.79 in the 92-node map. These restrictions also change the anatomical-division mix and do not, by themselves, separate mapping precision from division composition. **(B1)** Mean numbers of P545, C171, T245 and H350 family genes in each node's top-4,000 marker panel are shown by anatomical division, ordered by mean cascade reach. **(B2)** Across the 82 nodes, H350 marker-panel count is positively associated with the number of node-specific dimers (Spearman $\rho = +0.73$), whereas P545 count is negatively associated ($\rho = -0.42$). **(B3)** Rank correlations of node-by-node marker-panel, dimer, cascade-engagement and CORUM-complex similarities with mouse projection connectivity are shown before and after partialling inter-centroid distance. Dark bars include all 82 nodes; pale bars exclude five nodes whose cascades reached no proteins. Correlations were tested with 4,999 node-label permutations; "ns" denotes $p \geq 0.05$. Values of all bars not marked "ns" cleared the statistical significance threshold. The CORUM engaged-subunit-fraction profile has the highest distance-partialled correlation among the plotted products (+0.243 for 82 nodes; +0.287 for 77). Sets of complexes passing $q < 0.05$ (the $q < 0.05$ CORUM set arm) are nearly empty and show no detectable signal because a median of one complex per node clears BH correction. The cascade reaches 675 of 8,695 CNS-expressed proteins, so a complex needs nearly all its subunits within that set to pass BH over 1,132 tests. The CORUM complexes' half-engaged and engaged-fraction arms have the sufficient power to clear the statistical significance threshold. CORUM analyses use the 8,695-gene CNS-expressed background.

**Figure 7. Characterization of unique and shared contributions of the seven-axis association matrix defined in the 82-node CSDX model of human and mouse brain connectomes.** Panels designated **(a)** and **(b)** show each of seven individual predictor values for human AHBA CSDX model **(a)** and the original mouse ISH model **(b)**, respectively, against the unchanged 80-node mouse projection matrix from the 82-node CSDX model source. Each horizontal bar combines drop-one unique $R^2$ in the seven-axis model (colored) with its non-disjoint overlapping portion of the standalone predictor $R^2$ (light grey). Labels next to each bar give unique $R^2$, one-sided empirical node-label p from 4,999 permutations, and BH q corrected over 21 axis×map tests within each arm. Dashed lines indicate each species seven-axis fitted $R^2$. Panels designated **(c)** and **(d)** partition full fitted $R^2$ into seven unique terms and a net signed shared balance, the sum of all 2–7-way commonality terms, including negative suppression components. Human and original mouse data sets use 8,978 and 10,567 genes, respectively, so no cross-species comparisons of matched orthologues can be inferred from the reported fractions. Candidate PPI dimers and CORUM fractions were derived from expression, databases of PPI and protein complexes; no physical assembly measurements were analyzed. Additional human-mouse brain connectome model comparisons, including the seven-axis mouse–human predictor comparisons on matched orthologues, are reported in the Supplementary Information.

**Figure 8. Aggregate unique and net shared fractions of the matched seven-axis models.** In each 73/82/92-node source map, the 70/80/87 included nodes and 8,984/8,978/8,919 one-to-one orthologues are common to mouse ISH gene expression and human AHBA gene expression arms, both predicting the unchanged mouse projection outcome. Bars are the sum of seven conditional drop-one unique $R^2$ values divided by that arm's fitted full-model $R^2$; the complementary net shared fraction is full-model $R^2$ minus the unique sum, divided by full-model $R^2$, and includes signed commonality/suppression. Shown above the bars are human-minus-mouse percentage-point differences, basic 95% intervals from 1,999 paired node-weight resamples, and two-sided p/BH q for a family of nine primary map×aggregate contrasts. The right panel changes the CORUM feature to a ≥50%-engaged binary set. Assays and species-specific coordinate frames are different and the parcellations overlap; neither the proportions nor their contrast establish a biological species effect.

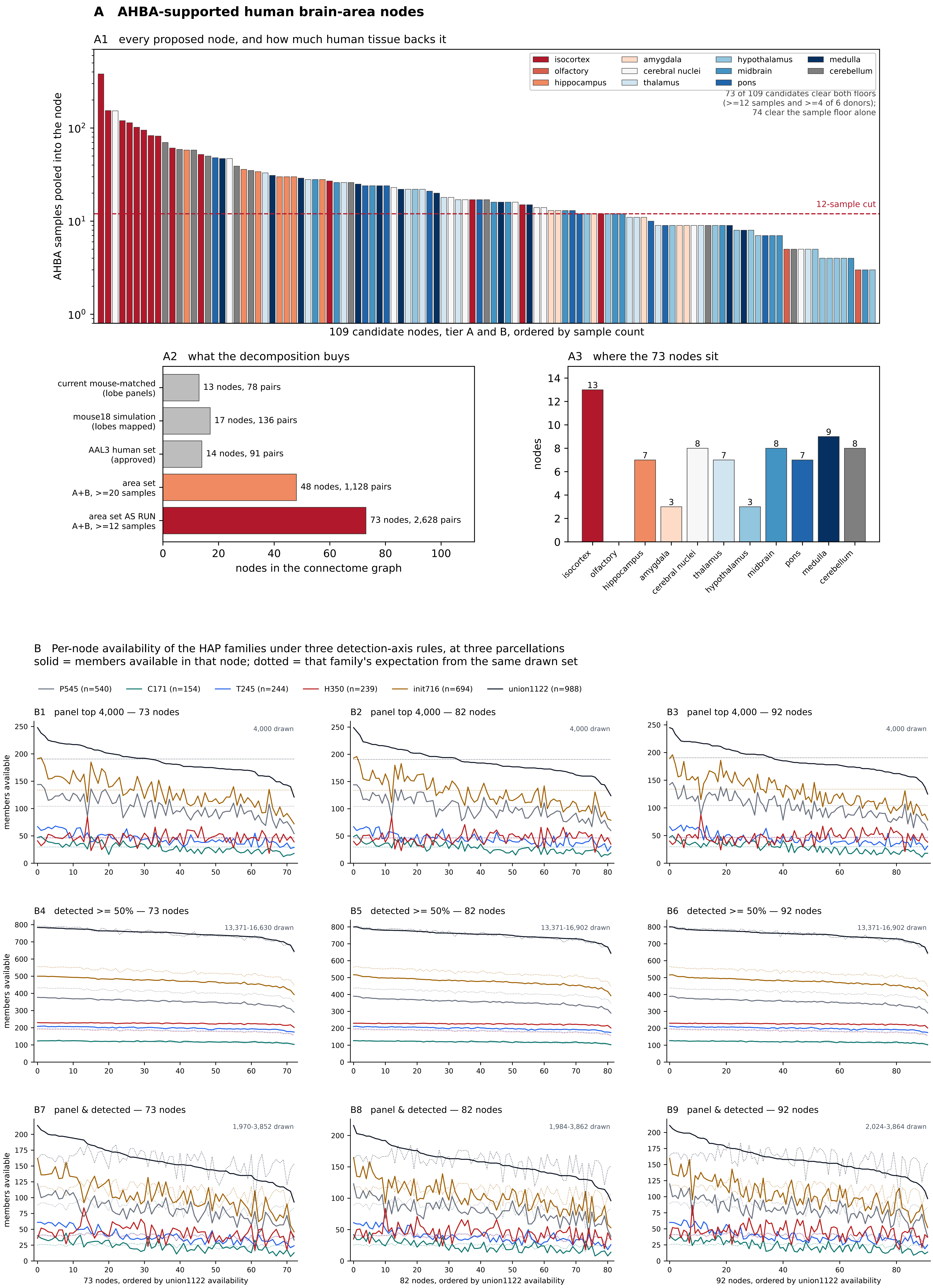

Availability is per-node DETECTION: detFrac[gene, node] is the fraction of that node's samples, pooled over every donor that sampled it, with a present AHBA PACall; a gene is available where detFrac >= 0.5. Population is the same 20,737 attempted genes in every panel; only the rule changes. Node order is set independently in each panel by union1122 availability, so x positions are not comparable across panels. Under 'panel top 4,000' the drawn set is fixed at 4,000 and the expectation is flat; under the two detection rules the drawn set varies node to node, which is why the dotted lines move.

## A HAP family availability on the DETECTION axis, 73-node connectome

Detection is called per node: the fraction of that node's samples — pooled over every donor that sampled it — in which PACall calls the gene present, thresholded at 50%.
The rule is absolute, so unlike the top-4,000 panel rule it permits a gene to be available in every node. Population for every test is the 20,737 genes attempted on the array;
each rule's drawn set is stated per node and the hypergeometric uses that node's own drawn size, BH across nodes within a family. The breadth-matched null in e and f replaces
each member with a non-member detected in exactly the same number of nodes, so only the anatomical placement of detection differs.

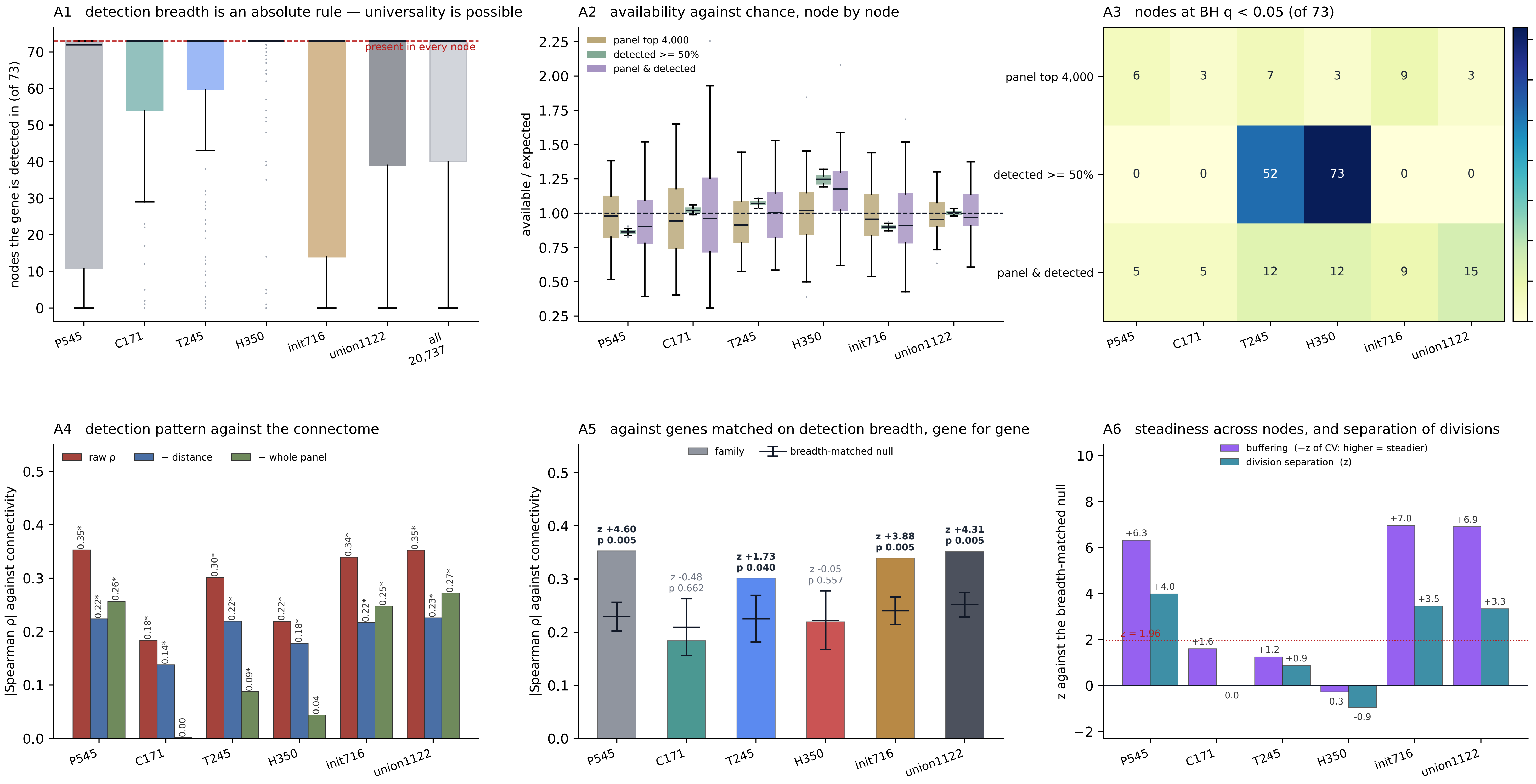

## B HAP family availability on the DETECTION axis, 82-node connectome

Detection is called per node: the fraction of that node's samples — pooled over every donor that sampled it — in which PACall calls the gene present, thresholded at 50%.
The rule is absolute, so unlike the top-4,000 panel rule it permits a gene to be available in every node. Population for every test is the 20,737 genes attempted on the array;
each rule's drawn set is stated per node and the hypergeometric uses that node's own drawn size, BH across nodes within a family. The breadth-matched null in e and f replaces
each member with a non-member detected in exactly the same number of nodes, so only the anatomical placement of detection differs.

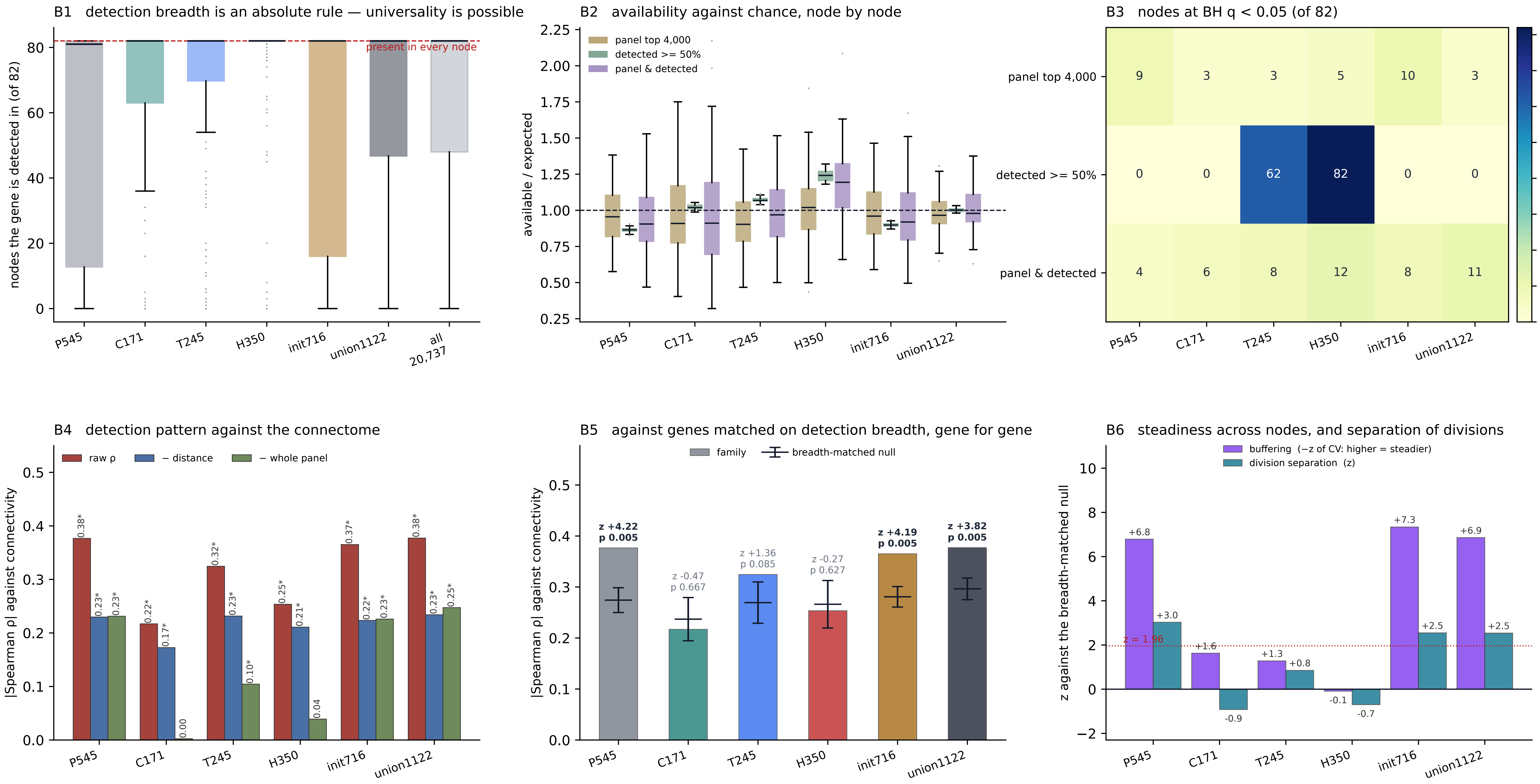

## C HAP family availability on the DETECTION axis, 92-node connectome

Detection is called per node: the fraction of that node's samples — pooled over every donor that sampled it — in which PACall calls the gene present, thresholded at 50%.
The rule is absolute, so unlike the top-4,000 panel rule it permits a gene to be available in every node. Population for every test is the 20,737 genes attempted on the array;
each rule's drawn set is stated per node and the hypergeometric uses that node's own drawn size, BH across nodes within a family. The breadth-matched null in e and f replaces
each member with a non-member detected in exactly the same number of nodes, so only the anatomical placement of detection differs.

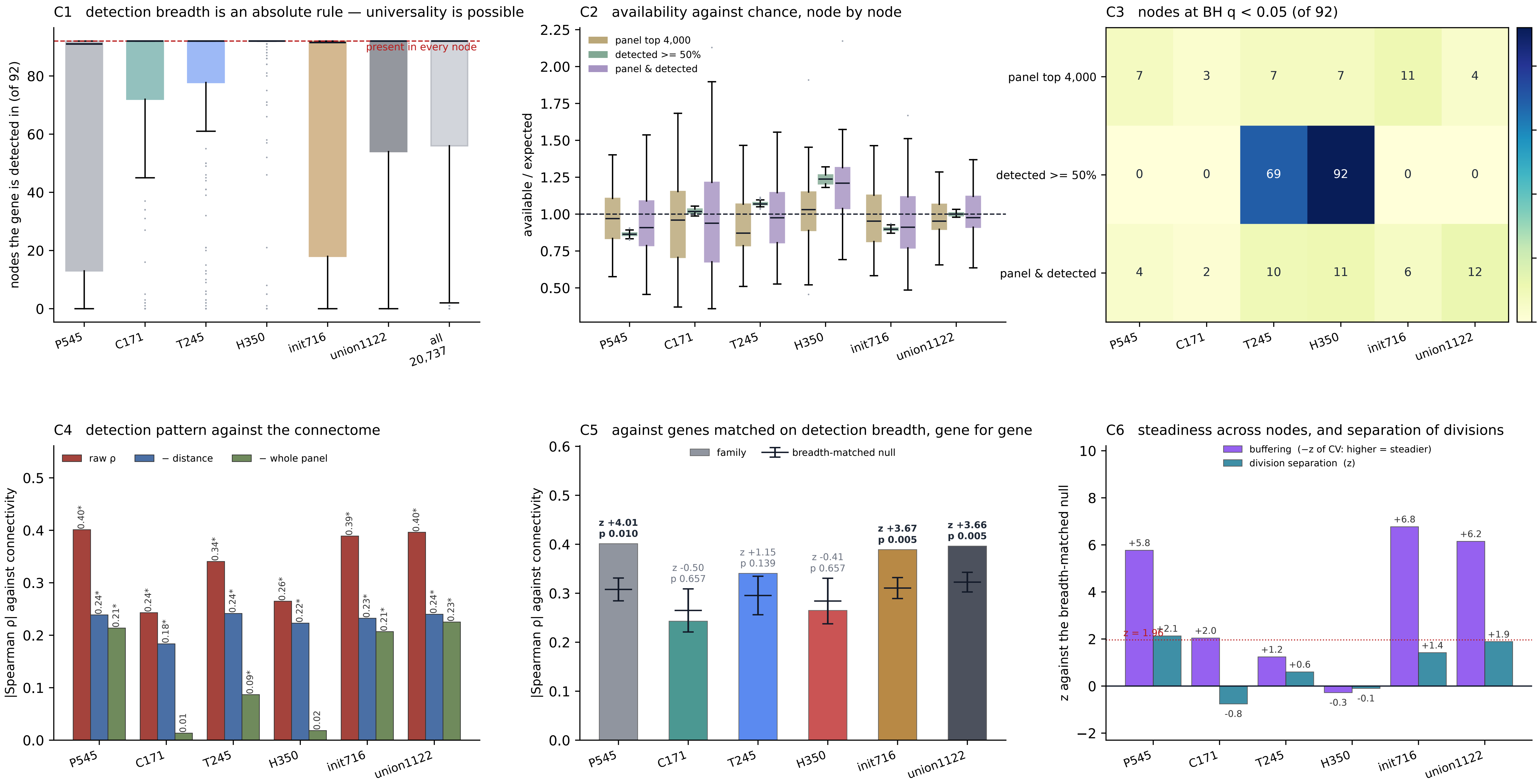

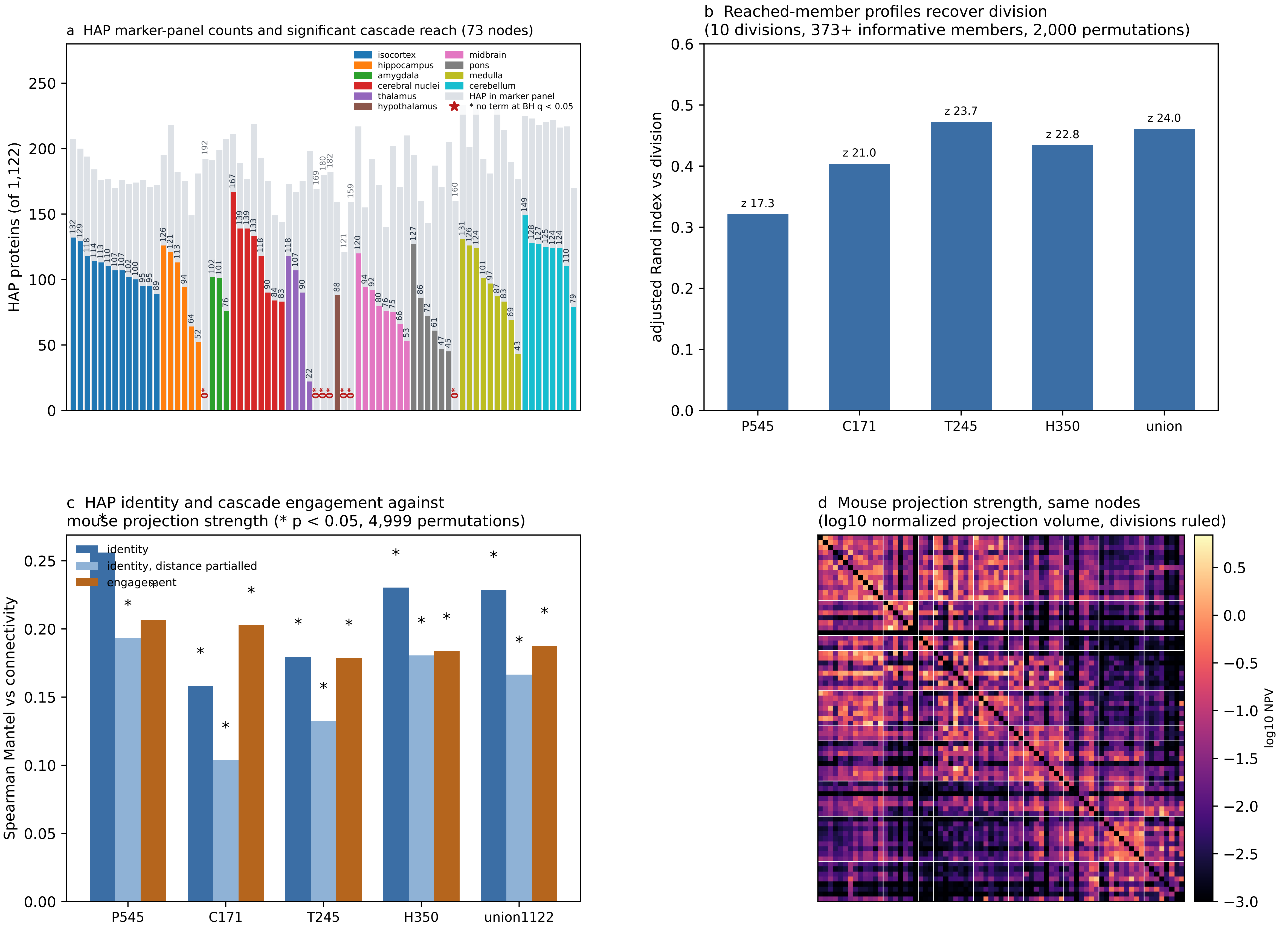
a HAP marker-panel counts and significant cascade reach (73 nodes)
HAP proteins (of 1,122)
isocortex
hippocampus
amygdala
cerebral nuclei
thalamus
hypothalamus
midbrain
pons
medulla
cerebellum
HAP in marker panel
* no term at BH q < 0.05
b Reached-member profiles recover division
(10 divisions, 373+ informative members, 2,000 permutations)
adjusted Rand index vs division
z 17.3
z 21.0
z 23.7
z 22.8
z 24.0
P545
C171
T245
H350
union
c HAP identity and cascade engagement against
mouse projection strength (* p < 0.05, 4,999 permutations)
Spearman Mantel vs connectivity
identity
identity, distance partialled
engagement
P545
C171
T245
H350
union1122
d Mouse projection strength, same nodes
(log10 normalized projection volume, divisions ruled)
log10 NPV

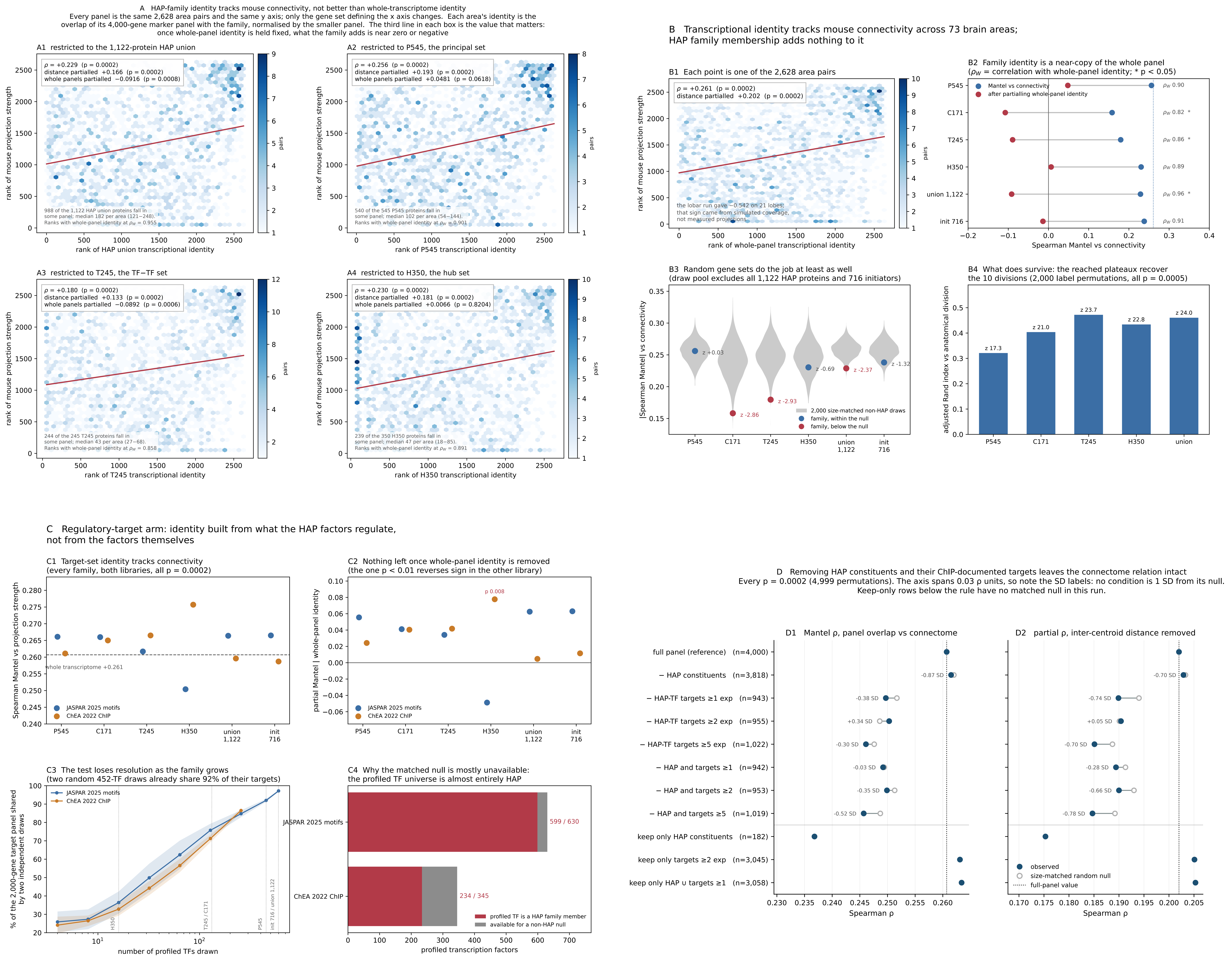

A HAP-family identity tracks mouse connectivity, not better than whole-transcriptome identity
Every panel is the same 2,628 area pairs and the same y axis; only the gene set defining the x axis changes. Each area's identity is the overlap of its 4,000-gene marker panel with the family, normalised by the smaller panel. The third line in each box is the value that matters: once whole-panel identity is held fixed, what the family adds is near zero or negative
A1 restricted to the 1,122-protein HAP union
ρ = +0.229 (p = 0.0002)
distance partialled +0.166 (p = 0.0002)
whole panels partialled −0.0916 (p = 0.0008)
988 of the 1,122 HAP union proteins fall in some panel; median 182 per area (121–248). Ranks with whole-panel identity at ρW = 0.955
rank of HAP union transcriptional identity
rank of mouse projection strength
pairs
A2 restricted to P545, the principal set
ρ = +0.256 (p = 0.0002)
distance partialled +0.193 (p = 0.0002)
whole panels partialled +0.0481 (p = 0.0618)
540 of the 545 P545 proteins fall in some panel; median 102 per area (54–144). Ranks with whole-panel identity at ρW = 0.901
rank of P545 transcriptional identity
A3 restricted to T245, the TF−TF set
ρ = +0.180 (p = 0.0002)
distance partialled +0.133 (p = 0.0002)
whole panels partialled −0.0892 (p = 0.0006)
244 of the 245 T245 proteins fall in some panel; median 43 per area (27–68). Ranks with whole-panel identity at ρW = 0.858
rank of T245 transcriptional identity
A4 restricted to H350, the hub set
ρ = +0.230 (p = 0.0002)
distance partialled +0.181 (p = 0.0002)
whole panels partialled +0.0066 (p = 0.8204)
239 of the 350 H350 proteins fall in some panel; median 47 per area (18–85). Ranks with whole-panel identity at ρW = 0.891
rank of H350 transcriptional identity
B Transcriptional identity tracks mouse connectivity across 73 brain areas; HAP family membership adds nothing to it
B1 Each point is one of the 2,628 area pairs
ρ = +0.261 (p = 0.0002)
distance partialled +0.202 (p = 0.0002)
the lobar run gave −0.542 on 21 lobes; that sign came from simulated coverage, not measured projections
rank of whole-panel transcriptional identity
rank of mouse projection strength
pairs
B2 Family identity is a near-copy of the whole panel (ρW = correlation with whole-panel identity; * p < 0.05)
Mantel vs connectivity
after partialling whole-panel identity
P545 ρW 0.90
C171 ρW 0.82 *
T245 ρW 0.86 *
H350 ρW 0.89
union 1,122 ρW 0.96 *
init 716 ρW 0.91
Spearman Mantel vs connectivity
B3 Random gene sets do the job at least as well (draw pool excludes all 1,122 HAP proteins and 716 initiators)
|Spearman Mantel| vs connectivity
z +0.03
z -2.86
z -2.93
z -0.69
z -2.37
z -1.32
2,000 size-matched non-HAP draws
family, within the null
family, below the null
P545
C171
T245
H350
union 1,122
init 716
B4 What does survive: the reached plateaux recover the 10 divisions (2,000 label permutations, all p = 0.0005)
adjusted Rand index vs anatomical division
z 17.3
z 21.0
z 23.7
z 22.8
z 24.0
P545
C171
T245
H350
union
C Regulatory-target arm: identity built from what the HAP factors regulate, not from the factors themselves
C1 Target-set identity tracks connectivity (every family, both libraries, all p = 0.0002)
Spearman Mantel vs projection strength
whole transcriptome +0.261
JASPAR 2025 motifs
ChEA 2022 ChIP
C2 Nothing left once whole-panel identity is removed (the one p < 0.01 reverses sign in the other library)
partial Mantel | whole-panel identity
p 0.008
C3 The test loses resolution as the family grows (two random 452-TF draws already share 92% of their targets)
% of the 2,000-gene target panel shared by two independent draws
number of profiled TFs drawn
H350
T245 / C171
P545
init 716 / union 1,122
C4 Why the matched null is mostly unavailable: the profiled TF universe is almost entirely HAP
JASPAR 2025 motifs
599 / 630
ChEA 2022 ChIP
234 / 345
profiled TF is a HAP family member
available for a non-HAP null
profiled transcription factors
D Removing HAP constituents and their ChIP-documented targets leaves the connectome relation intact
Every p = 0.0002 (4,999 permutations). The axis spans 0.03 ρ units, so note the SD labels: no condition is 1 SD from its null. Keep-only rows below the rule have no matched null in this run.
D1 Mantel ρ, panel overlap vs connectome
D2 partial ρ, inter-centroid distance removed
full panel (reference) (n=4,000)
− HAP constituents (n=3,818) -0.87 SD -0.70 SD
− HAP-TF targets ≥1 exp (n=943) -0.38 SD -0.74 SD
− HAP-TF targets ≥2 exp (n=955) +0.34 SD +0.05 SD
− HAP-TF targets ≥5 exp (n=1,022) -0.30 SD -0.70 SD
− HAP and targets ≥1 (n=942) -0.03 SD -0.28 SD
− HAP and targets ≥2 (n=953) -0.35 SD -0.66 SD
− HAP and targets ≥5 (n=1,019) -0.52 SD -0.78 SD
keep only HAP constituents (n=182)
keep only targets ≥2 exp (n=3,045)
keep only HAP ∪ targets ≥1 (n=3,058)
observed
size-matched random null
full-panel value
Spearman ρ

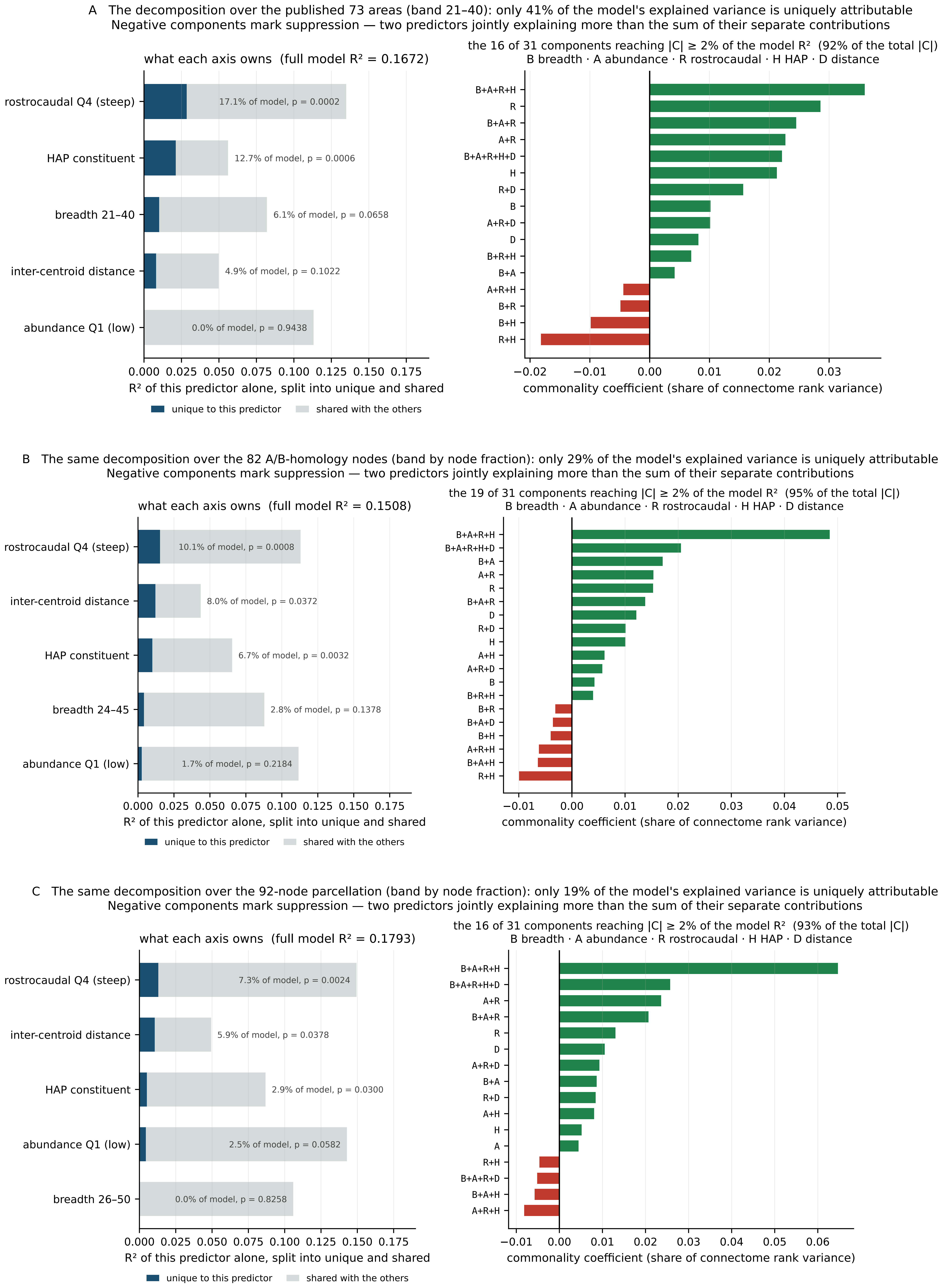
A The decomposition over the published 73 areas (band 21–40): only 41% of the model's explained variance is uniquely attributable
Negative components mark suppression — two predictors jointly explaining more than the sum of their separate contributions
what each axis owns (full model R² = 0.1672)
rostrocaudal Q4 (steep)
17.1% of model, p = 0.0002
HAP constituent
12.7% of model, p = 0.0006
breadth 21–40
6.1% of model, p = 0.0658
inter-centroid distance
4.9% of model, p = 0.1022
abundance Q1 (low)
0.0% of model, p = 0.9438
0.000 0.025 0.050 0.075 0.100 0.125 0.150 0.175
R² of this predictor alone, split into unique and shared
unique to this predictor
shared with the others
the 16 of 31 components reaching |C| ≥ 2% of the model R² (92% of the total |C|)
B breadth · A abundance · R rostrocaudal · H HAP · D distance
B+A+R+H
R
B+A+R
A+R
B+A+R+H+D
H
R+D
B
A+R+D
D
B+R+H
B+A
A+R+H
B+R
B+H
R+H
−0.02 −0.01 0.00 0.01 0.02 0.03
commonality coefficient (share of connectome rank variance)
B The same decomposition over the 82 A/B-homology nodes (band by node fraction): only 29% of the model's explained variance is uniquely attributable
Negative components mark suppression — two predictors jointly explaining more than the sum of their separate contributions
what each axis owns (full model R² = 0.1508)
rostrocaudal Q4 (steep)
10.1% of model, p = 0.0008
inter-centroid distance
8.0% of model, p = 0.0372
HAP constituent
6.7% of model, p = 0.0032
breadth 24–45
2.8% of model, p = 0.1378
abundance Q1 (low)
1.7% of model, p = 0.2184
0.000 0.025 0.050 0.075 0.100 0.125 0.150 0.175
R² of this predictor alone, split into unique and shared
unique to this predictor
shared with the others
the 19 of 31 components reaching |C| ≥ 2% of the model R² (95% of the total |C|)
B breadth · A abundance · R rostrocaudal · H HAP · D distance
B+A+R+H
B+A+R+H+D
B+A
A+R
R
B+A+R
D
R+D
H
A+H
A+R+D
B
B+R+H
B+R
B+A+D
B+H
A+R+H
B+A+H
R+H
−0.01 0.00 0.01 0.02 0.03 0.04 0.05
commonality coefficient (share of connectome rank variance)
C The same decomposition over the 92-node parcellation (band by node fraction): only 19% of the model's explained variance is uniquely attributable
Negative components mark suppression — two predictors jointly explaining more than the sum of their separate contributions
what each axis owns (full model R² = 0.1793)
rostrocaudal Q4 (steep)
7.3% of model, p = 0.0024
inter-centroid distance
5.9% of model, p = 0.0378
HAP constituent
2.9% of model, p = 0.0300
abundance Q1 (low)
2.5% of model, p = 0.0582
breadth 26–50
0.0% of model, p = 0.8258
0.000 0.025 0.050 0.075 0.100 0.125 0.150 0.175
R² of this predictor alone, split into unique and shared
unique to this predictor
shared with the others
the 16 of 31 components reaching |C| ≥ 2% of the model R² (93% of the total |C|)
B breadth · A abundance · R rostrocaudal · H HAP · D distance
B+A+R+H
B+A+R+H+D
A+R
B+A+R
R
D
A+R+D
B+A
R+D
A+H
H
A
R+H
B+A+R+D
B+A+H
A+R+H
−0.01 0.00 0.01 0.02 0.03 0.04 0.05 0.06
commonality coefficient (share of connectome rank variance)

A   Restricting to precisely remapped nodes raises the division separation, on fewer nodes
Filled markers mark permutation p < 0.05; same statistic and the same breadth-matched 200-draw null in all six arms

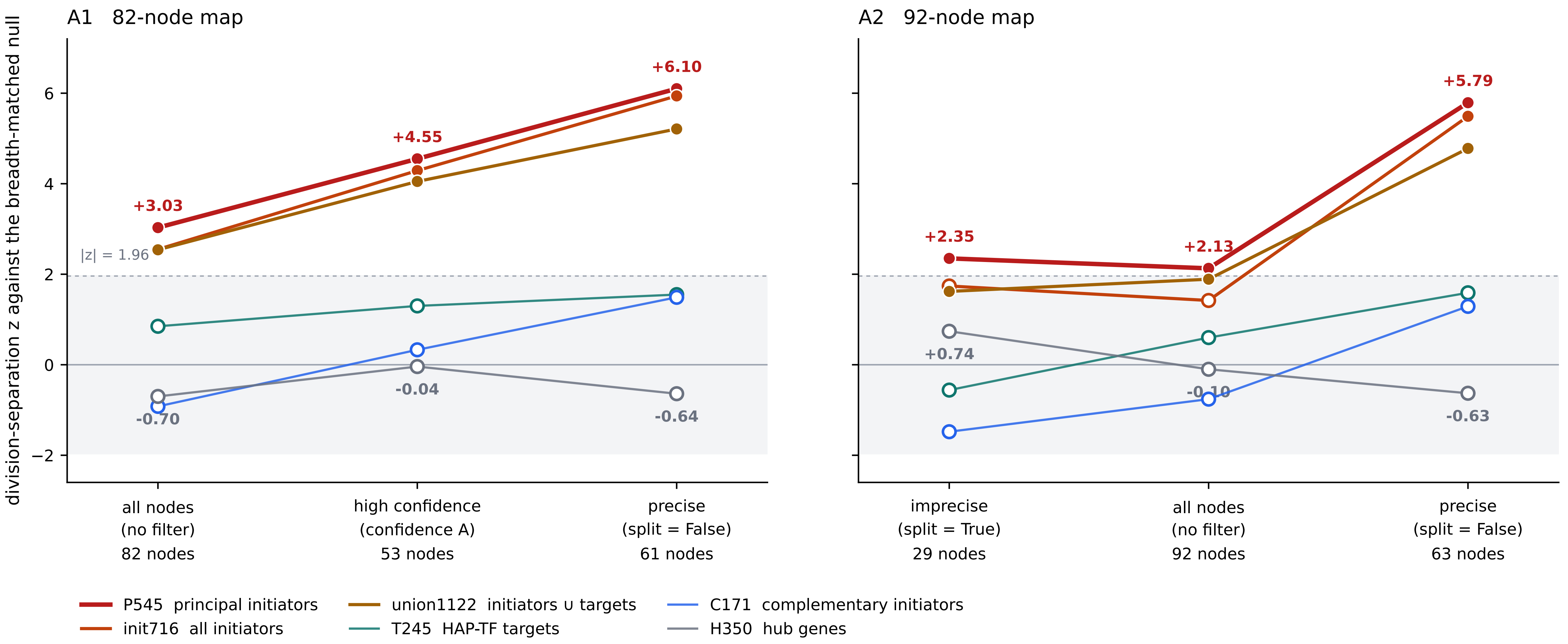

Dropping the 21 loosely mapped nodes from the 82-node set doubles P545 (+3.03 → +6.10) on fewer node pairs, which is the opposite of what scale dilution predicts.  H350 does not move on either axis.  Definitions, the same in both panels: PRECISE = split = False, one AHBA structure group went to one mouse target undivided; IMPRECISE = split = True, the node was built by dividing a parent structure group; HIGH CONFIDENCE = the 82-node map's own A/B grade.  Precision and confidence are different criteria, so a's three arms are not nested.
Filtering on remap precision also rebalances divisions, because the split nodes are mostly isocortical — 82-precise keeps 5 of 22 isocortical nodes and 92-imprecise is 23 of 29 isocortical across 4 divisions — so precise-against-imprecise is partly cortical-against-subcortical, and these arms do not adjudicate scale against mapping noise on their own.

B   Node-specific dimers and CORUM complexes both track the connectome, and the complex-level profile tracks it best
82-node A/B-homology map; per-node top-4,000 marker panel drives the cascade, CORUM tested on the 8,695-gene CNS-expressed background

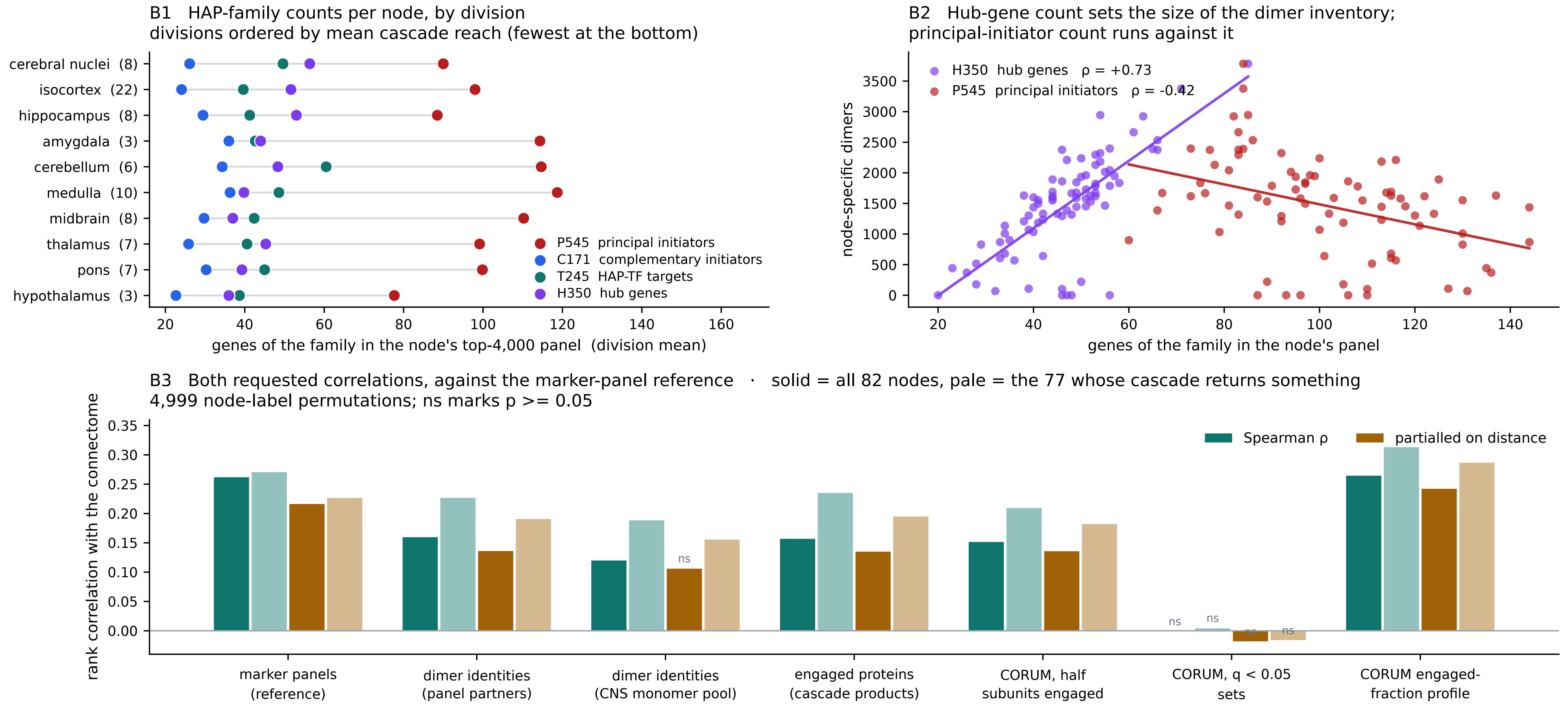

Every arm is a node-by-node similarity matrix correlated with the connectome over the upper triangle, so all of them descend from the same marker panels — the question is what survives the cascade, not whether the arms are independent.  The engaged-fraction profile over all 1,132 CORUM complexes is the strongest arm once distance is removed (+0.243 on 82 nodes, +0.287 on 77), slightly above the panel it descends from.  The q < 0.05 CORUM sets carry no signal because they are nearly empty: a median of 1 complex per node clears BH, since the cascade reaches 675 of 8,695 CNS-expressed proteins and a complex needs almost all its subunits inside that set to pass.  Five nodes reach nothing at all (CA4/hilus, mediodorsal thalamus, lateral thalamus ventral, lateral hypothalamic area, pontine nuclei); an all-empty row is Jaccard 0 against every other node, a similarity the cascade never measured, so the pale bars drop those five.

## 82-node source: seven-axis shared and unique fractions

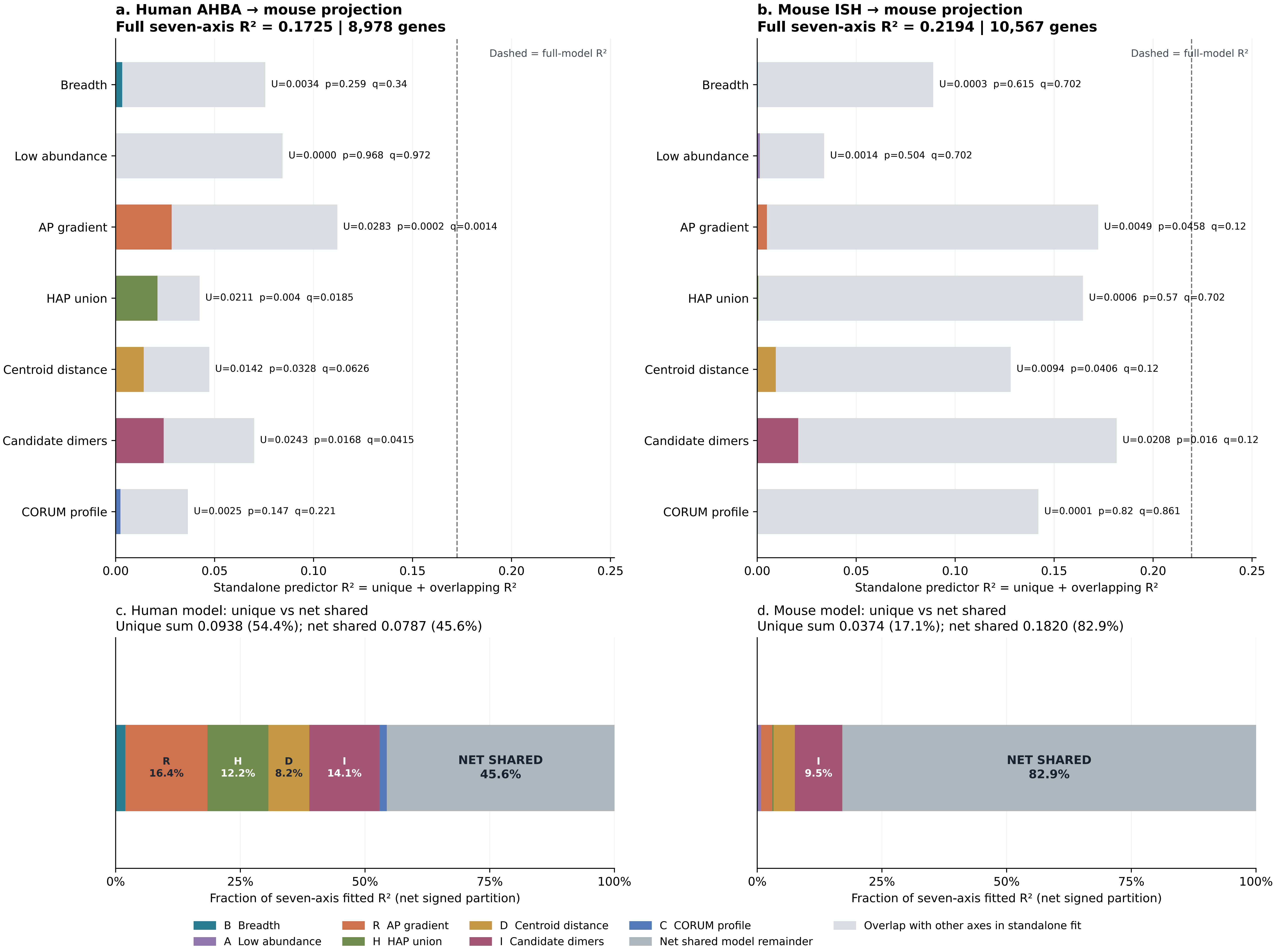

Original mouse: 10,567 genes. Earlier reference human AHBA: 8,978 joint orthologs. Both use 80 nodes; gene universes differ, so do not test species differences from these fractions.

Net shared is the SIGNED sum of 2–7-way commonality (includes negative components); pale grey per-axis overlaps are NOT disjoint model pieces. q = BH within each source arm over 21 map×axis tests.

**Matched unique versus net shared contributions**

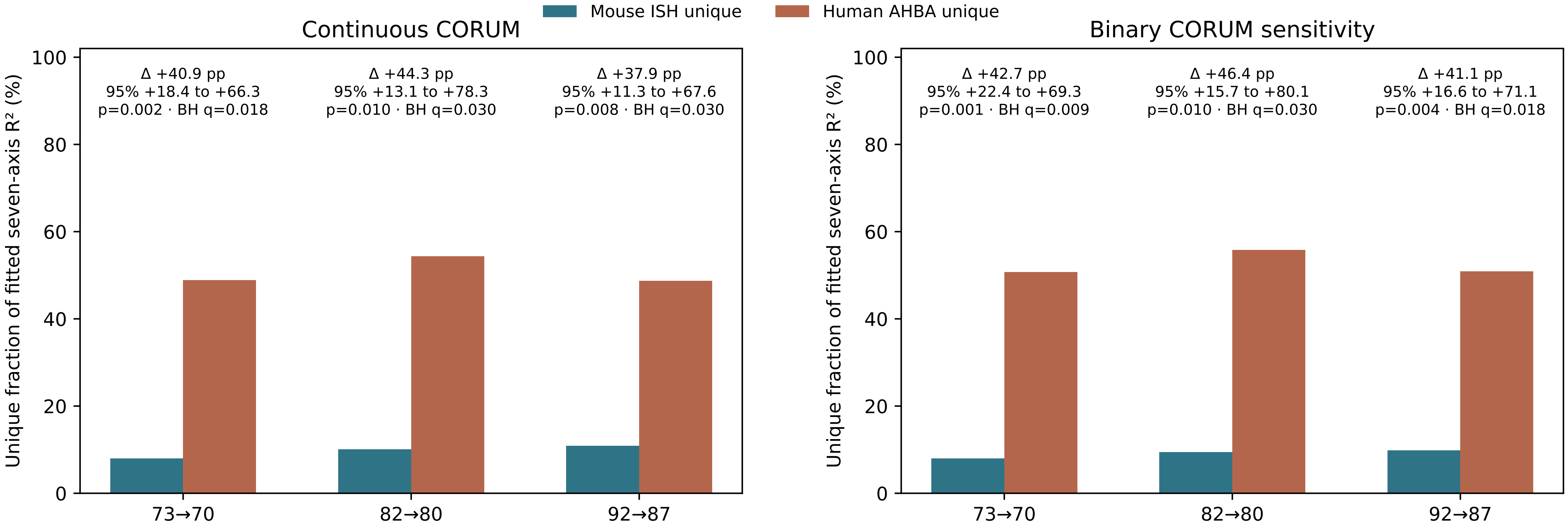

Net shared = full fitted $R^2$ − sum of seven drop-one unique $R^2$; includes signed commonality/suppression.

Intervals: paired node-weight basic bootstrap, 1,999 draws. ISH/AHBA assays and species spatial coordinates differ; maps overlap.